\documentclass[12pt]{article}
\usepackage{palatino,epsfig,makeidx,amsfonts,amsmath}
\usepackage{amsthm}
\usepackage[round]{natbib}
\usepackage{rotating}
\usepackage[usenames]{color}

\def\E{\mbox{E}}

\def\N{\mbox{N}}

\def\Be{\mbox{Be}}

\def\I{\mbox{I}}

\newtheorem{lemma}{Lemma}
\newtheorem{theorem}{Theorem}

\begin{document}

\title{Individual adaptation: an adaptive MCMC scheme for variable selection problems}

\author{J. E. Griffin, K. {\L}atuszy\'nski and M. F. J. Steel\thanks{Jim Griffin is Professor, School of Mathematics, Statistics and Actuarial Science, University of Kent, Canterbury, CT2 7NF, U.K. (Email: J.E.Griffin-28@kent.ac.uk), Krys {\L}atuszy\'nski is Royal Society University Research Fellow and Assistant Professor (Email: K.G.Latuszynski@warwick.ac.uk) and Mark Steel is Professor, Department of Statistics, University of Warwick, Coventry, CV4 7AL, U.K. (Email: m.steel@warwick.ac.uk). The authors are grateful to B{\l}a\.zej Miasojedow for helpful comments.}}

\date{}
\maketitle

\abstract{The increasing size of data sets has lead to variable selection in regression becoming increasingly important. Bayesian approaches are attractive since they allow uncertainty about the choice of variables to be formally included in the analysis. The application of fully Bayesian variable selection methods to large data sets is computationally challenging. We describe an adaptive Markov chain Monte Carlo approach called {\it Individual Adaptation} which adjusts a general proposal to the data. We show that the algorithm is ergodic and discuss its use within parallel tempering and sequential Monte Carlo approaches. We illustrate the use of the method on two data sets including a gene expression analysis with 22 577 variables.}


\section{Introduction}

The problem of choosing a subset of potential variables to include in a linear model is an important, and well-studied, problem in statistics. Let $y$ be an $(n\times 1)$-dimensional vector of responses and $X$ be an $(n\times p)$-dimensional data matrix.
The indicator variable $\gamma_i$ denotes whether the $i$-th variable is included in the model (when $\gamma_i=1$) and we define $p_{\gamma}=\sum_{j=1}^p \gamma_j$. The linear regression model is
\[
y=\alpha {\bf 1} + X_{\gamma}\beta_{\gamma} + \epsilon
\]
where ${\bf 1}$ is an $(n\times 1)$-dimensional vector of 1's, $X_{\gamma}$ is the sub-matrix of $X$ where the $i$-th column is included if $\gamma_i=1$, $\beta_{\gamma}$ is a $(p_{\gamma}\times 1)$-dimensional vector and $\epsilon\sim\N(0,\sigma^2 I_n)$. It will be useful to define the notation $\theta_{\gamma}=(\alpha,\beta_{\gamma})$.

Bayesian methods are attractive for the variable selection problem since they can formally incorporate uncertainty about the form of the model and provide Bayesian model averaged (BMA) estimates of common parameters and predictions.
 These can be substantially more accurate than those from a single model.
A prior distribution is placed on the parameters $\theta_{\gamma}$ and $\sigma^2$ jointly with the model $\gamma$.  The most commonly used prior structure is
\begin{equation}p(\alpha,\sigma^2,\beta_{\gamma},\gamma)\propto \sigma^{-2}p(\beta_{\gamma}\vert\sigma^2, \gamma)p(\gamma)\label{prior}
\end{equation}
\begin{eqnarray*} \textrm{with} \qquad \beta_{\gamma}  \vert  \sigma^2,\gamma & \sim & \N(0,\sigma^2 V_{\gamma}) \qquad \textrm{and} \qquad p(\gamma)  =  h^{p_{\gamma}}(1-h)^{p-p_{\gamma}}.\end{eqnarray*}
The hyperparameter $0<h<1$ is the prior probability that a particular variable is included in the model and $V_{\gamma}$ is often chosen as proportional to $(X_\gamma^T X_\gamma)^{-1}$ (a $g$-prior) or to an identity matrix (implying conditional prior independence between the regression coefficients).
 The use of these methods extends beyond regression problems and underlies Bayesian approaches to many problems, such as flexible curve and surface estimation.

Posterior inference is challenging since the number of models ($2^p$) is very large if $p$ is not small and the
posterior distribution may be highly multi-modal. Interest normally centres around low-dimensional summaries such as posterior inclusion probabilities (PIP's) or predictive distributions for future observations.
There is  a large literature on computational strategies for model uncertainty problems and, particularly, regression models, see e.g. \cite{george1997approaches, dellaportas2002bayesian, o2009review, BoRi10, ClGhLi11} and references therein.
There are two main computational approaches:  Markov chain Monte Carlo (MCMC) sampling and heuristic search methods aiming to find the highest posterior probability models. \cite{GDMB13} provide an interesting comparison of these two methods which they term empirical and renormalization respectively. They show that the renormalization method is prone to biased estimates of posterior probabilities whereas the MCMC method can provide consistent estimates. Successful estimation using the empirical method depends on having a representative sample from the posterior distribution. This is challenging since the model space is large and the posterior distribution is potentially multi-modal. Many MCMC schemes have been proposed for this model  \citep[see {\it e.g.}][]{GDMB13} but these increasingly struggle to provide representative samples as $p$ becomes larger.
The difficulty of sampling from the posterior distribution
 is a particular problem with large numbers of covariates which is becoming increasingly common in many applications (with $p$ in the tens of thousands).

The complexity of the posterior distribution has lead to interest in methods where the computational algorithm adapts to the data. For example, \cite{Kwon11} consider building transition probabilities using the correlation matrix of the regressors.
 Alternatively, the algorithm can be adapted during the run. \cite{nottkohn05} developed a Gibbs sampling algorithm which allows the algorithm to adapt to the marginal inclusion probabilities (the posterior probability that a variable is included in the model). \cite{richardson2010bayesian} focus on high dimensional sparse multi-response regression models, that are central to genomics, and develop an adaptive Gibbs sampler for identifying hot spots in this context. \cite{lamnisos12} construct a tuneable proposal distribution in a Metropolis-Hastings algorithm and describe an adaptive algorithm which tunes this parameter to achieve a pre-specified average acceptance rate. \cite{JiSchmidler} use a mixture distribution for the proposal kernel and  adapt its parameters to minimize the Kullback-Leibler divergence from the target distribution.
  The problem of multi-modality can be addressed using standard computational techniques such as parallel tempering or sequential Monte Carlo samplers \citep[][with application to variable selection]{ScCh11} which use powered versions of the posterior distribution.

This paper describes  a flexible adaptive Metropolis-Hastings algorithm that is cheap to implement per iteration and is able to efficiently traverse the model space.
 This leads to substantially more efficient algorithms than commonly-used methods.
The adaptation step relies on the optimal acceptance rate criterion \citep{RGG97,roberts1998optimal, RoRo01}.
The adaptation parameter is a vector of length $2p$ which allows the deletion and addition of each variable conditional on the current model to be optimised individually. This flexibility allows the variables included in the model to change quickly and leads to substantial improvements in mixing. Each individual adaptation step is cheap as the marginal likelihood is calculated using a fraction of the variables which has the same order as the typical {\it a posteriori} model size. We also show
how this adaptive kernel can be used as a building block for interchain adaptation, parallel tempering and sequential Monte Carlo schemes in more challenging multi-modal problems. We also verify its ergodicity under some typical regularity assumptions.

The paper is organised as follows: Section \ref{sec:indiv_adap} introduces a new adaptive kernel for variable selection which we term ``individual adaptation'', Section \ref{sec:RAPA} discusses some methods for accelerating the convergence of the algorithm to the target acceptance probability. Section \ref{par_temp} considers their use as a building block in more complex algorithms for exploring posteriors with well-separated modes. Ergodicity of the algorithms is discussed in Section \ref{sec:ergodcity}. Section \ref{sec:numericals} presents the application of the methods to datasets with $p=100$ and $p=22\ 576$ possible covariates, and Section \ref{sec:concl} concludes. Supplementary material includes proofs of the ergodicity of the algorithms and a further example using sequential Monte Carlo and parallel tempering methods. Matlab code is available from \\\verb+http://www.kent.ac.uk/smsas/personal/jeg28/index.htm+.

\section{The individual adaptation algorithm} \label{sec:indiv_adap}

We will consider inference in Bayesian variable selection with a linear regression model and conjugate prior as in (\ref{prior}) using a Metropolis-Hastings sampler. In this case, the marginal likelihood $p(y\vert \gamma)$ can be calculated analytically and a sampler can be directly run on $\gamma$.


We define  a very general proposal on model space with parameters $A=(A_1,\dots,A_p)$, $D=(D_1,\dots,D_p)$ with $0<A_j,D_j<1$ and $\eta=(A, D)$.
A new model, $\gamma'$, is proposed independently, conditional on $\gamma$, according to the transition density
\[
q_{\eta}(\gamma,\gamma')=p(\gamma'\vert \gamma) = \prod_{j=1}^p p(\gamma'_j\vert \gamma_j)=\prod_{j=1}^p q_{\eta,j}(\gamma_j,\gamma'_j)
\] where 
$
q_{\eta,j}(\gamma_j=0,\gamma'_j=1)=A_j$, $q_{\eta,j}(\gamma_j=0,\gamma_j'=0)=1-A_j$,
$q_{\eta,j}(\gamma_j=1,\gamma'_j=0)=D_j$, and $q_{\eta,j}(\gamma_j=1,\gamma_j'=1)=1-D_j$.
The values of $\gamma'_1,\dots,\gamma'_p$ are conditionally independent and so can be quickly sampled.
The tuning parameter $A_j$ is the probability that the $j$-th variable is added to the model (if it is currently excluded)
and $D_j$ is the probability that the $j$-th variable is deleted from the model (if it is currently included).
 The proposed model is accepted using the standard Metropolis-Hastings acceptance probability
\[
a_{\eta}(\gamma,\gamma') = \min\left\{1,\frac{p(y\vert \gamma')p(\gamma')q_{\eta}(\gamma',\gamma)}{p(y\vert \gamma)p(\gamma)q_{\eta}(\gamma,\gamma')}\right\}.
\]
The proposal allows multiple variables to be added or deleted from the model and, consequently, we do not need separate add, remove or swap moves as in the standard multi-move proposal
\citep{BVF98}. If the number of additions and deletions is different, the model size will be proposed to change.
The expected proposed change in the model size, given $\gamma$, is
$ \sum_{i=1}^p \I(\gamma_j=0)A_j- \sum_{i=1}^p \I(\gamma_j=1)D_j$
and the total number of variables proposed to be changed is
$ \sum_{i=1}^p \I(\gamma_j=0)A_j+ \sum_{i=1}^p \I(\gamma_j=1)D_j$.
Unconditionally, these equal $\sum_{i=1}^p p(\gamma_j=0\vert y) A_j-\sum_{i=1}^p p(\gamma_j=1\vert  y)D_j $ and
$\sum_{i=1}^p p(\gamma_j=0\vert y) A_j+\sum_{i=1}^p p(\gamma_j=1\vert  y)D_j $ respectively.
Therefore, smaller values of $A_j$ and $D_j$ will tend to lead to smaller changes in the model. However, the effect on proposed model size of changing an individual $A_j$ or $D_j$ depends on the posterior inclusion probability (PIP) for the $j$-th variable. The value of $A_j$ will only have a large effect on the average size of change if $p(\gamma_j=0\vert y)$ is large.
The proposal is more general than the one proposed by \cite{lamnisos09} and is easily extended to allow the probabilities of adding or deleting each variable from the model to change over the run of the sampler.

Working with this proposal seems, at first, problematic since  there are $2p$ tuning parameters $A$ and $D$ which must be specified at the start of the algorithm and we have little guidance on their choice. Our solution is to follow the idea of \cite{lamnisos12} and choose values of these tuning parameters which give a pre-specified acceptance rate by adapting these tuning parameters during the MCMC run.
\cite{ScCh11} note that the usual form of average acceptance rate for Metropolis-Hastings samplers is not appropriate for the variable selection problem (or other problems on discrete spaces) where moves which do not change the model ({\it i.e.}~$\gamma$ and $\gamma'$ are the same) have positive probability. These have an acceptance probability of 1 but do not help mixing since the model does not change. They suggest using instead the  mutation rate which is defined to be \
\begin{equation}
\bar{a}_M=\int C(\gamma,\gamma')a_\eta(\gamma,\gamma')q_{\eta}(\gamma,\gamma')p(\gamma\vert y) d\gamma'\,d\gamma,\label{SA1}
\end{equation}
where $C(\gamma,\gamma')=0$ if $\gamma'_j=\gamma_j$  for all $j$ and $1$ otherwise.

The {\bf individual adaptation (IA) algorithm} targets a particular value, $\tau$, of the mutation rate.
Let $\gamma^{(i)}$ be the value of $\gamma$ at the start of the $i$-th iteration, $\gamma'$
be the subsequently proposed value and $\eta^{(i)}=(A^{(i)},D^{(i)})$ be the value of the tuning parameters used at the $i$-th iteration. We define for $j=1,\dots,p$
 \[
\gamma^{A\,(i)}_j=
\left\{\begin{array}{ll}
1\mbox{ if }\gamma'_j\neq \gamma_j^{(i)}\mbox{ and }\gamma_j^{(i)}=0\\
0\mbox{ otherwise}
\end{array}\right.
\]
\[
\gamma^{D\,(i)}_j=
\left\{\begin{array}{ll}
1\mbox{ if }\gamma'_j\neq \gamma_j^{(i)}\mbox{ and }\gamma_j^{(i)}=1\\
0\mbox{ otherwise}
\end{array}\right.
\]
The values of $A^{(i)}$ and $D^{(i)}$ are adapted using for $j=1\dots,p$ 
\footnotesize
\begin{equation} \label{eqn:adap_A}
\log\left(\frac{A^{(i+1)}_j-\epsilon}{1-A^{(i+1)}_j-\epsilon}\right)=\log\left(\frac{A^{(i)}_j-\epsilon}{1-A^{(i)}_j-\epsilon}\right)+
\phi_i\,\gamma^{A\,(i)}_j
\left(a_{\eta^{(i)}}\left(\gamma^{(i)},\gamma'\right) - \tau\right)
\end{equation} \normalsize
and \footnotesize
\begin{equation} \label{eqn:adap_B}
\log\left(\frac{D^{(i+1)}_j-\epsilon}{1-D^{(i+1)}_j-\epsilon}\right)=\log\left(\frac{D^{(i)}_j-\epsilon}{1-D^{(i)}_j-\epsilon}\right)+
\phi_i\,\gamma^{D\, (i)}_j
\left(a_{\eta^{(i)}}\left(\gamma^{(i)},\gamma'\right) - \tau\right)
\end{equation} \normalsize
where 
$0<\epsilon<1/2$ and $\epsilon$ is small,
$\phi_i=O(i^{-\lambda})$ for some constant $1/2<\lambda\leq 1$ and
$a_{\eta^{(i)}}\left(\gamma^{(i)},\gamma'\right)$ represents the acceptance
probability at the $i$-th iteration. The transformation implies that $\epsilon<A_j^{(i)}<1-\epsilon$ and
$\epsilon<D_j^{(i)}<1-\epsilon$ and the algorithm targets an average mutation rate of $\tau$ if that is attainable.
Clearly, if the current acceptance probability exceeds $\tau$, $A_j$ for the currently excluded variables will be increased, as well as $D_j$ for the variables that are in the current model. This implies larger proposed model changes, so will tend to decrease the mutation rate.

The starting values of $A$ and $D$ can have a considerable effect on the convergence of the tuning parameters towards values which have an average mutation rate of $\tau$. We have found that the following starting values work well in practice:
$
A^{(1)}_j={\nu}/\{(1-h)p\}
$
and
$
D^{(1)}_j={\nu}/(h p),
$
where $h p$ is the prior mean model size (see (\ref{prior})). The range of values taken by $A^{(1)}_j$ and $D^{(1)}_j$ imply that $\epsilon (1-h)p<\nu<(1-\epsilon) hp$ if $h<1/2$ (which will be true in large $p$ settings).
If the initial value of $\gamma$ is generated from the prior, this choice of $A^{(1)}$ and $D^{(1)}$ implies that the expected number of proposed changes from $\gamma^{(1)}$ is $2\nu$.  We have used the value $\nu=1$ in our examples and found that the
 performance of the algorithm is robust to choices in the range $0.25$ to $4$ in Example 1.

The efficiency of the algorithm with respect to the choice of $\tau$ has been empirically studied in Example 1 and appears not to be very sensitive as long as $\tau$ is not too close to $0$ or $1$. This confirms other empirical and theoretical studies on scaling and in particular we note that for discrete state spaces the optimal $\tau$ will depend on the problem, see e.g. Figure 3 on page 282 of \cite{roberts1998optimal}. The accelerated versions of the algorithm described below (RAPA and MCA) appear even more robust to the choice of $\tau.$

\section{Accelerated individual adaptation algorithms} \label{sec:RAPA}

The convergence of $A$ and $D$ can be slow if $p$ is large. This does not affect the ergodicity of the adaptive chain but it can affect the mixing of the chain in MCMC  runs of practically sensible length. Therefore, we consider two possible methods for accelerating the algorithm. The first method uses $r$ independent MCMC chains but shares the proposal parameters across the chains
which are updated after the iteration of each independent chain.
\cite{craiu2009learn} empirically show that a related approach improves the rate of convergence of adaptive algorithms towards their target acceptance rate in the context of the classical Adaptive Metropolis algorithm of \cite{haario2001adaptive} (see also Bornn et al. 2013)
\nocite{BJDD12}. This will be referred to as {\bf multiple chain acceleration (MCA)} (which differs from the parallel tempering methods described in Section~\ref{par_temp}).

A second approach uses the {\bf reverse acceptance probability acceleration (RAPA)} method.
The individual adaptation algorithm updates $A_j$ only if $\gamma^{A\,(i)}_j=1$ or $D_j$ only if $\gamma^{D\,(i)}_j=1$. This potentially wastes information since the Metropolis-Hastings algorithm considers a pair of models (the current and the proposed) and we only use the acceptance probability for moving from the
current to the proposed. The Metropolis-Hastings acceptance ratio for the reverse move from
proposed to current, $a_{\eta}(\gamma',\gamma)$, can also be calculated using the values needed to compute $a_{\eta}(\gamma,\gamma')$. To include the acceptance probability $a_{\eta}(\gamma',\gamma)$ in the update of $A$ and $D$, we need to keep the mutation rate targeting $\tau$ in the
stochastic approximation algorithm. This is $\bar{a}_M$ in (\ref{SA1}),
which is just an expectation with respect to the posterior and the transition. A second chain $(\delta,\delta')$ can be constructed from $\gamma$ and $\gamma'$ in the following way
\[
(\delta,\delta')=\left\{\begin{array}{ll}
(\gamma',\gamma) & \mbox{with probability }a_{\eta}(\gamma,\gamma')\\
(\gamma,\gamma') & \mbox{with probability }1-a_{\eta}(\gamma,\gamma')\\
\end{array}\right..
\]
The stationary distribution of $\delta$ is the posterior distribution due to properties of the Metropolis-Hastings algorithm
which implies  that
$p(\delta,\delta')=p(\delta\vert y)q_{\eta}(\delta,\delta')$. It follows that we can write (\ref{SA1}) as
\begin{align*}
&\int C(\delta,\delta')'a_{\eta}(\delta,\delta')q_{\eta}(\delta,\delta')p(\delta\vert y)\,d\delta'\,d\delta\\
=&\E[C(\delta,\delta')a_{\eta}(\delta,\delta')]\\
=&a_{\eta}(\gamma,\gamma')\E[C(\gamma',\gamma)a_{\eta}(\gamma',\gamma)]
+\left(1-a_{\eta}(\gamma,\gamma')\right)\E[C(\gamma,\gamma')a_{\eta}(\gamma,\gamma')].
\end{align*}
Taking a weighted average of this expression and  $\E[C(\gamma,\gamma')a_{\eta}(\gamma,\gamma')]$  gives
\[
wa_{\eta}(\gamma,\gamma')\E[C(\gamma',\gamma)a_{\eta}(\gamma',\gamma)]
+\left(1-wa_{\eta}(\gamma,\gamma')\right)\E[C(\gamma,\gamma')a_{\eta}(\gamma,\gamma')].
\]

Therefore, and noticing that $C(\gamma,\gamma')=C(\gamma',\gamma)$, an accelerated version of
the adaptive algorithm (the {\bf IA-RAPA algorithm}) uses the following updates:\tiny
\begin{eqnarray} \label{eqn:adap_A_1}
\log\left(\frac{A^{(i+1)}_j-\epsilon}{1-A^{(i+1)}_j-\epsilon}\right)
&= & \log\left(\frac{A^{(i)}_j-\epsilon}{1-A^{(i)}_j-\epsilon}\right)+\phi_i\,
\gamma^{A\,(i)}_j
\left(a_{\eta^{(i)}}\left(\gamma^{(i)},\gamma'\right) - \tau\right) \left(1-w a_{\eta^{(i)}}\left(\gamma^{(i)},\gamma'\right)\right),
\\ \label{eqn:adap_A_2}
\log\left(\frac{D^{(i+1)}_j-\epsilon}{1-D^{(i+1)}_j-\epsilon}\right)
&=&
\log\left(\frac{D^{(i)}_j-\epsilon}{1-D^{(i)}_j-\epsilon}\right)+\phi_i\,\gamma^{A\,(i)}_j\left(a_{\eta^{(i)}}\left(\gamma',\gamma^{(i)}\right) - \tau\right)  wa\left(\gamma^{(i)},\gamma'\right),
\\ \label{eqn:adap_D_1}
\log\left(\frac{D^{(i+1)}_j-\epsilon}{1-D^{(i+1)}_j-\epsilon}\right)
&=&
\log\left(\frac{D^{(i)}_j-\epsilon}{1-D^{(i)}_j-\epsilon}\right)+\phi_i\,\gamma^{D\,(i)}_j\left(a_{\eta^{(i)}}\left(\gamma^{(i)},\gamma'\right) - \tau\right) \left(1 -w a_{\eta^{(i)}}\left(\gamma^{(i)},\gamma'\right)\right)
\\ \label{eqn:adap_D_2}
\log\left(\frac{A^{(i+1)}_j-\epsilon}{1-A^{(i+1)}_j-\epsilon}\right),
& = &
\log\left(\frac{A^{(i)}_j-\epsilon}{1-A^{(i)}_j-\epsilon}\right)+\phi_i\,\gamma^{D\,(i)}_j\left(a_{\eta^{(i)}}\left(\gamma',\gamma^{(i)}\right) - \tau\right) wa_{\eta^{(i)}}\left(\gamma^{(i)},\gamma'\right).
\end{eqnarray}\normalsize

Whereas $w=0$ corresponds to the standard IA algorithm, we will use $w=0.5$ for IA-RAPA in the applications below.

\section{Multi-modal posterior distributions}
\label{par_temp}

The individual adaptation algorithm behaves like a Metropolis-Hastings random walk (albeit running on a very high-dimensional space). However, in common with all random walk samplers, the algorithm can become stuck in local modes if the
modes are sufficiently well-separated.
We will consider methods which use a sequence of annealed versions of the posterior distribution
\[
\pi_k(\gamma \vert y)\propto p(y\vert \gamma)^{t_k} \pi(\gamma),\qquad  k=1,\dots,m
\]
where the parameters $0<t_1<t_2<\dots<t_m=1$ are referred to as temperatures (with smaller $t_j$ referring to higher temperatures).
The density $\pi_m(\gamma\vert y)$ is the posterior density $p(\gamma\vert y)$ of interest.
The density at other temperatures  will be flatter than the posterior distribution and is more likely to allow for moves between the local modes.
Our adaptive algorithm is potentially well-suited to this approach since it can quickly explore the model space at high temperatures (the posterior raised to a power close to 0) and so rapidly move between local modes.
We consider two implementations: a parallel tempering and a sequential Monte Carlo algorithm \citep{ScCh11}  which use a sequence of annealed versions of the posterior distribution.

The {\bf parallel tempering (PT) algorithm} has long been used to improve convergence of MCMC algorithms for multi-modal posterior distributions. Its use in MCMC for Bayesian variable selection was first proposed by \cite{JaStHo07}.
The algorithm runs a chain at each temperature and proposes to swap the current value in two chains in such a way that the chains are drawn from the correct distribution.
The idea is formalized by defining  a joint target 
for $\gamma^{\star}=(\gamma^{\star}_1,\dots,\gamma^{\star}_m)$,
\[
\pi(\gamma^{\star}\vert y)=\prod_{k=1}^m \pi_k(\gamma^{\star}_k\vert y)
\]
 where $\pi_k(\gamma^{\star}_k\vert y)\propto
p(y\vert \gamma^{\star}_k)^{t_k} \pi(\gamma^{\star}_k)$.
 An MCMC algorithm is run on the target $\pi(\gamma^{\star}\vert y)$ with two types of moves. Firstly, an MCMC algorithm  updates $\gamma^{\star}_k$ for all values of $k$. Secondly, a Metropolis-Hastings algorithm is introduced which proposes to swap $\gamma^{\star}_k$ with $\gamma^{\star}_l$ where $k$ and $l$ are drawn from some distribution. In practice, the proposed value is often chosen by first drawing a value $k$ uniformly from $\{1,\dots,m-1\}$ and then choosing $l=k+1$. 
 This restricts the algorithm to swaps between chains at consecutive temperatures.

There are a number of drawbacks with this algorithm which can be addressed using adaptive ideas. Firstly, the temperature schedule $t_1,\dots,t_{m-1}$ must be chosen. Recent work has suggested that the optimal choice of temperature schedule should maintain an acceptance rate of 0.234 for swaps between chains \citep{AtRoRo11}. An adaptive algorithm that exploits this idea is suggested by \cite{MiMoVi12} and adopted in our algorithm. Secondly, the distribution for higher temperatures (smaller values of $t_j$) should be relatively flat to allow easier exploration. However,
standard variable selection algorithms may move slowly across these targets since only one variable is changed in the model at each iteration. We use different tuning parameters for each chain and define $\eta_k$ to be the value of the tuning parameters for the $k$-th chain.
 The  individual adaptation algorithm allows more than one variable to be changed at each iteration in any chain and so should
 avoid the problem with standard variable selection algorithms. In summary, one iteration of the full {\bf individual adaptation-parallel tempering (IA-PT) algorithm} is
\begin{itemize}

\item For $k=1,\dots,m$ do  individual adaptation updating with $\pi_k$ as the target distribution and tuning parameters $\eta_k$.

\item Choose $k$ uniformly from $\{1,\dots,m-1\}$ and set $l=k+1$. Propose to swap $\gamma^{(k)}$ with $\gamma^{(l)}$ and accept the move with acceptance probability
\[
\min\left\{1,\frac{p\left(y\vert \gamma^{\star}_l\right)^{t_k} p\left(y\vert \gamma^{\star}_k\right)^{t_l}
}{p\left(y\vert \gamma^{\star}_k\right)^{t_k} p\left(y\vert \gamma^{\star}_l\right)^{t_l}}
\right\}.
\]

\item Let
$\rho^{(h)}_{j-1} = t^{(h)}_j-t^{(h)}_{j-1}, j=1,2,\dots,m-1$.
These values are updated to
\[
\rho^{(h+1)}_j =
\left\{\begin{array}{ll}
\rho^{(h)}_j  &\mbox{if }  j = 1,\dots,l-1,l+1,\dots,m-1,\\
 \rho^{(h)}_j + \zeta_h (a - \hat{a})&\mbox{if } j=l
\end{array}\right.
\]
where $\zeta_h$ is $O(h^{-\lambda})$ for some constant $1/2<\lambda\leq 1$, $a$ is the Metropolis-Hastings acceptance probability and $\hat{a}$ is the target average acceptance probability for the parallel tempering moves.
Finally, the temperatures are updated to
$
t^{(h+1)}_j = t^{(h+1)}_{j-1} + \rho^{(h+1)}_{j-1},$
$j=1,2,\dots,m-1.
$
\end{itemize}

As we discussed in Section 3, multiple chains can lead to faster convergence of the proposal parameters. A multiple chain acceleration version of the IA-PT algorithm can be defined where all chains share the same proposal parameters and temperature schedule and which will be referred to as the {\bf MCA-IA-PT algorithm}.


\cite{ScCh11} propose a related {\bf sequential Monte Carlo (SMC) algorithm} using the sequence of distributions
$\pi_1(\gamma\vert y),\dots,\pi_m(\gamma\vert y)$.
They suggest sampling from this sequence of distribution using an SMC algorithm and choosing the sequence of powers $t_j$ adaptively.
The {\bf IA-SMC algorithm} proceeds by alternating selection steps with MCMC steps as follows. Let $t_0=0$ and $N$ particles
$\gamma^{\dagger}_1,\dots,\gamma^{\dagger}_N$  are chosen from $\pi_0(\gamma^{\dagger}_i)=\pi(\gamma^{\dagger}_j)$.
\begin{enumerate}

\item At the $k$-th selection step - calculate the weight of the $j$-th particle which is distributed according to $\pi_{k-1}$ as
\[
w_j\propto  p\left(y\left\vert \gamma^{\dagger}_j\right.\right)^{t_k-t_{k-1}},\qquad j=1,\dots,N.
\]
A sample which reweights according to $w_1,\dots,w_N$ is selected. Any reweighting scheme can be used but we have used systematic resampling in our examples. The new sample is distributed according to $\pi_k$. The value of $t_k$ is chosen so that the Effective Sample Size is approximately $c N$ for some $0<c<1$.

\item MCMC step - $K$ iterations of the individual adaptation algorithm are run for each particle using a common set of $A$ and $D$.

\end{enumerate}

The algorithm proceeds until $t_k=1$. We have chosen the value $c=0.9$. This is a conservative choice and often leads to small changes from $t_{k-1}$ to $t_k$ but smaller values of $c$ typically lead to substantially increased problems with particle degeneracy. This leads to a value of $m$ which is chosen adaptively and so  is random.
The individual  adaptation algorithm for each $k$ starts from the values of $A$ and $D$ at the end of the $(k-1)$-th step but the iteration counter is re-set. This allows the algorithm to use information about these tuning parameters from updating the chains for $\pi_1,\dots,\pi_{k-1}$ but also allows these values to be quickly adapted at each step.
The tuning parameters are assumed common for all particles and so changes in the shape from $\pi_{k-1}(\gamma\vert y)$ to $\pi_k(\gamma\vert y)$ can be quickly learnt in the algorithm. An alternative scheme for adaptation in SMC is discussed by \cite{FeTa12}.

\section{Ergodicity of the Algorithms} \label{sec:ergodcity}

Since adaptive MCMC algorithms violate the Markov condition, the standard and well developed Markov chain theory can not be used to establish ergodicity and we need to derive appropriate results for our algorithms. In particular, it is well known that even simple and seemingly reasonable adaptive algorithms may fail to converge \citep{MR2340211, bai2011containment, latuszynski2013adaptive}.

Here we provide some fairly general ergodicity results in the case when the model parameters can be integrated out and the marginal likelihood $p(y|\gamma)$ is available analytically.

Recall that $\pi(\gamma\vert y) \propto  p(y|\gamma)p(\gamma),$ the target posterior on the model space $M$ and the vector of adaptive parameters \[ \eta^{(i)} = (A^{(i)},D^{(i)}) \;\; \in \;\; [\varepsilon, 1-\varepsilon]^{2p} \; \equiv \; \Delta_{\varepsilon}\] at time $i.$ By $P_{\eta}(\gamma, \cdot)$ denote the non-adaptive Markov chain kernel corresponding to the fixed choice of $\eta.$ Thus under dynamics of the individual adaptation algorithm
\[
\mathbb{P}\Big[\gamma^{(i+1)} \in S \, \Big| \, \gamma^{(i)}= \gamma, \eta^{(i)} =\eta \Big] \; =  \; P_{\eta}(\gamma, S),
\qquad S \subseteq M.
\]
In the case of multiple chain acceleration, where $r$ copies of the chain are run, the respective model state space is the product space and thus the current state of the algorithm at time $i$  is $\gamma^{\otimes r,\, (i)} \in M^r$ and the stationary distribution is the product density $\pi^{\otimes r}$ on $M^r$. Clearly, when $r=1$ then the multiple chain becomes a single chain and thus all the notions and results in the sequel stated for multiple chains acceleration are valid for the single chain algorithm.

To assess ergodicity, we need to define the distribution of the adaptive algorithm at time $i$, and the associated total variation distance: for $S \subseteq M^{ r}$
\begin{eqnarray*}
\mathcal{L}^{(i)}\big[(\gamma^{\otimes r}, \eta), S \big] & := & \mathbb{P}\Big[\gamma^{\otimes r,\, (i)} \in S \, \Big| \, \Gamma_{0}= \gamma^{\otimes r}, \eta^{(0)} =\eta \Big], \\
T(\gamma^{\otimes r}, \eta,i) & := & \| \mathcal{L}^{(i)}\big[(\gamma^{\otimes r}, \eta), \cdot \big] - \pi^{\otimes r}(\cdot) \|_{TV} \\ & = & \sup_{S \in M^{ r}}|\mathcal{L}^{(i)}\big[(\gamma^{\otimes r}, \eta), S \big] - \pi^{\otimes r}(S)|.
\end{eqnarray*}
Defining $\pi(f)=\sum f(\gamma)\pi(\gamma\vert y)$, we show that all algorithms are ergodic, {\it i.e.}
\begin{eqnarray}
  \label{eq_thm:IA_erg}
\lim_{i \to \infty} T(\gamma^{\otimes r}, \eta, i) = 0 & \quad \textrm{for every } & \gamma^{\otimes r} \in M^{ r}, \quad 
\end{eqnarray}
and satisfy a Weak Law of Large Numbers, {\it i.e.}
\begin{eqnarray}
  \label{eq_thm:IA_WLLN}
{1 \over i}\sum_{k=1}^{i} f(\gamma_k) \; \stackrel {i\to\infty}{\longrightarrow} \; \pi(f) & \textrm{in probability,}&  \textrm{for every} \quad f:  M^{ r} \to \mathbb{R} \quad \\ & \quad \textrm{and every}& \gamma^{\otimes r, \,(0)} \in M^{ r}, \quad  \eta^{(0)} \in \Delta_{\varepsilon}. \nonumber
\end{eqnarray}
We first establish the following result.

\begin{lemma}\label{lem:SUE}
The kernel $P_{\eta}(\gamma,S)$ leads to a simultaneously uniform ergodic chain. For all $\delta > 0$ there exists $N=N(\delta) \in \mathbb{N}$ such that
\[
\|P_{\eta}^{N}(\gamma^{\otimes r}, \cdot) - \pi^{\otimes r}(\cdot)\|_{TV} \leq \delta \quad \textrm{for all} \; \gamma^{\otimes r} \in M^{ r} \; \textrm{and} \; \eta \in \Delta_{\varepsilon},
\]
\end{lemma}

Our first result considers non-tempered versions of the algorithm.
\begin{theorem}\label{thm:IA-PT_erg} Assume that $ p(y|\gamma)\pi(\gamma)$ is available analytically for all $\gamma\in M$ and $\varepsilon > 0$  in \eqref{eqn:adap_A}, \eqref{eqn:adap_B}, or in \eqref{eqn:adap_A_1}-\eqref{eqn:adap_D_2}, respectively. Then each of the algorithms: IA, RAPA-IA, MCA-IA and MCA-RAPA-IA is ergodic
and satisfies a Weak Law of Large Numbers.

\end{theorem}

A comprehensive analysis of the individual adaptation algorithm with other generalised linear models or with linear models whose parameters are given a non-conjugate prior distributions requires an involved case by case treatment, and is beyond the scope of this paper. However, we note that if the prior distributions are supported on a compact set and all involved densities are continuous and everywhere positive, establishing ergodicity for a specific model will, with some technical care, typically be possible.  The following theorem establish the ergodicity of the parallel tempered MCMC algorithm.
\begin{theorem}\label{thm:IA_erg} Assume that $p(y|\gamma)^t\pi(\gamma)$ is available analytically and is finite for all $0<t\leq 1$ and $\gamma\in M$ and $\varepsilon > 0$ in \eqref{eqn:adap_A}, \eqref{eqn:adap_B}, or in \eqref{eqn:adap_A_1}-\eqref{eqn:adap_D_2}, respectively. Then each of the algorithms: IA-PT, RAPA-IA-PT, MCA-IA-PT and MCA-RAPA-IA-PT is ergodic
and satisfies a Weak Law of Large Numbers.
\end{theorem}

Finally Theorem 1 combined with standard results for SMC algorithms can be used to show that the IA-SMC algorithm is ergodic as well as its variations with MCA and RAPA.

\section{Applications}\label{sec:numericals}

\subsection{Tecator Data}

The tecator data contains  172  observations and 100 variables. They have been previously analysed using Bayesian linear regression techniques by \cite{gribro10}, who give a description of the data, and
\cite{lamnisos12}. The prior used was (\ref{prior}) with $V_\gamma=100 I$ and $h=5/100$.
We generated 10 independent runs of the algorithms with different tuning parameters 
and without thinning. If multiple chain acceleration was used, the number of iterations in each chain was divided by the number of chains. This fixes the total number of iterations so that run times are the same for all algorithms.

\begin{figure}[h!]
\begin{center}
\includegraphics[trim=10mm 0mm 60mm 180mm, clip]{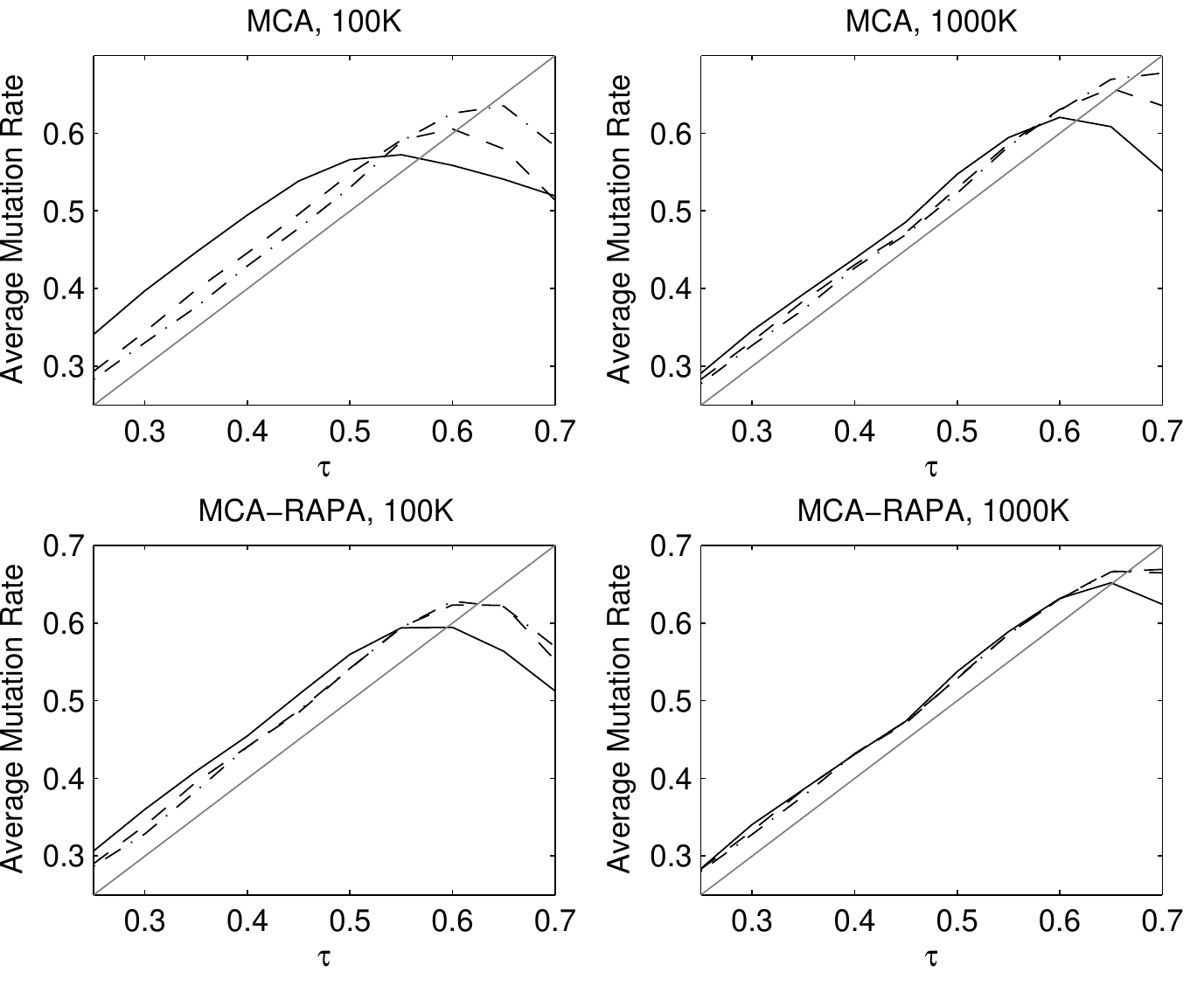}
\end{center}
\caption{\small Tecator data: average mutation rate  over 10 independent runs as a function of $\tau$ with
MCA only and MCA with RAPA 100\ 000 iterations and
 1\ 000\ 000 iterations after a burn-in of 100\ 000 iterations. The number of chains were: 1 (solid line), 5 (dashed line) and 25 (dot-dashed line). The thin solid line is $y=x$
}
\label{fig:tecator1_accept2}
\end{figure}
Figure~\ref{fig:tecator1_accept2} shows the average mutation rate as a function of $\tau$
for the IA algorithm with MCA only and IA-MCA with IA-RAPA. Both algorithm were able  to effectively target the chosen
average mutation rate 
for most values of $\tau$ with both 100\ 000 and 1\ 000\ 00 iterations after a burnin of 100\ 000 iterations.
Unsurprisingly, the targeting improves as the number of iterations or the number of chains is  increased. All algorithms struggle with targeting larger values of $\tau$ but these are not in a range that we would consider to be optimal.

\begin{figure}[h!]
\begin{center}
\includegraphics[trim=0mm 0mm 60mm 230mm, clip]{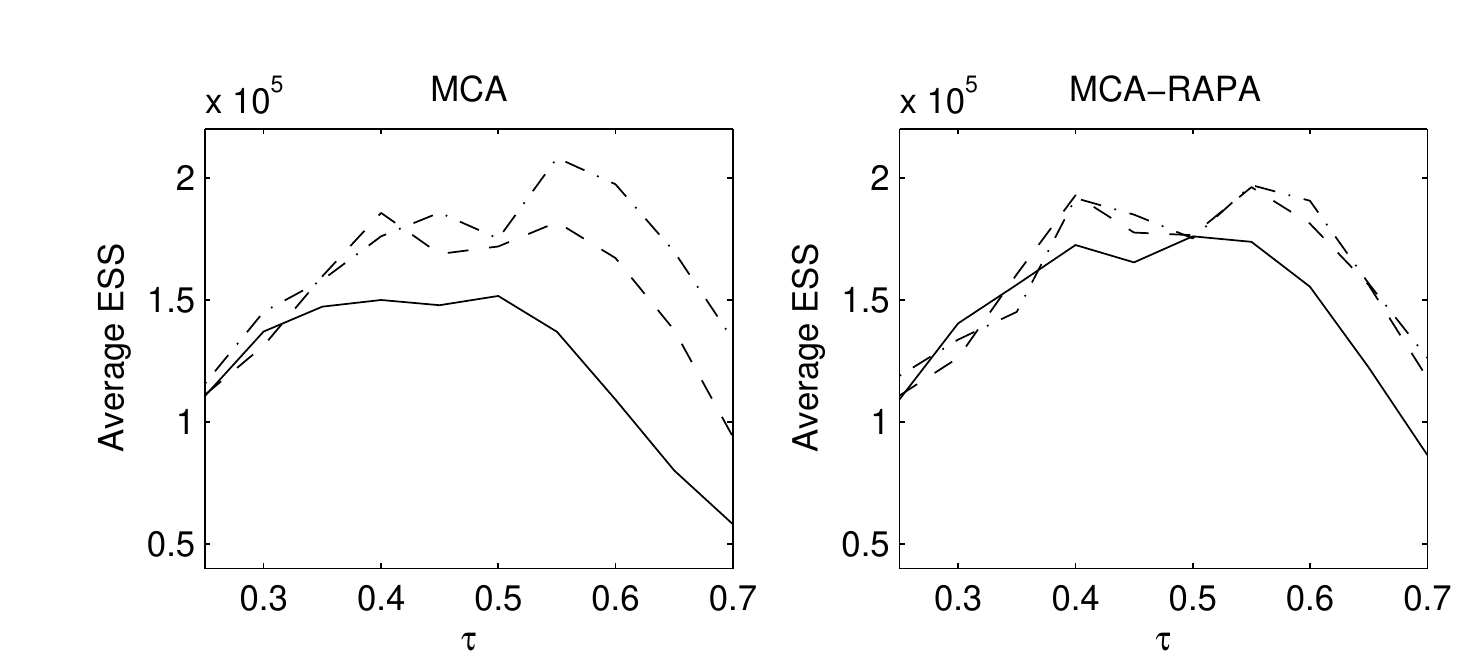}
\end{center}
\caption{\small Tecator data: average ESS over 10 independent runs as a function of $\tau$ using MCA only and MCA-RAPA.
The number of chains in MCA were: 1 (solid line), 5 (dashed line) and 25 (dot-dashed line)}
\label{fig:tecator1_ESS1}
\end{figure}
Figure~\ref{fig:tecator1_ESS1} shows the effect of $\tau$ on the average effective sample size (ESS) with different number of multiple chains
and with or without the RAPA step (using $w=0.5$).
 In all case, the ESS was maximized by $\tau$ between 0.35 and 0.55 but was relatively constant over this range.
 This is largely in keeping with previous work on optimal acceptance rates for Metropolis-Hastings random walk samplers on discrete spaces and implies that the performance of the algorithm is not overly sensitive to choice of $\tau$.
Both acceleration steps tended to lead to larger effective sample sizes at all values of $\tau$. The effect of MCA was much less pronounced when RAPA was used.
The improvement of MCA-RAPA over MCA in targeting
the correct rate (particularly, for a single chain) leads to the slightly larger ESS with the addition of a RAPA step.
\begin{figure}[h!]
\begin{center}
\begin{tabular}{ccc}
 IA-RAPA, $\tau=0.45$ & Lamnisos {\it et al}, $\tau=0.3$ & Multi-move MH\\
\includegraphics[trim=0mm 60mm 0mm 80mm, scale=0.25, clip]{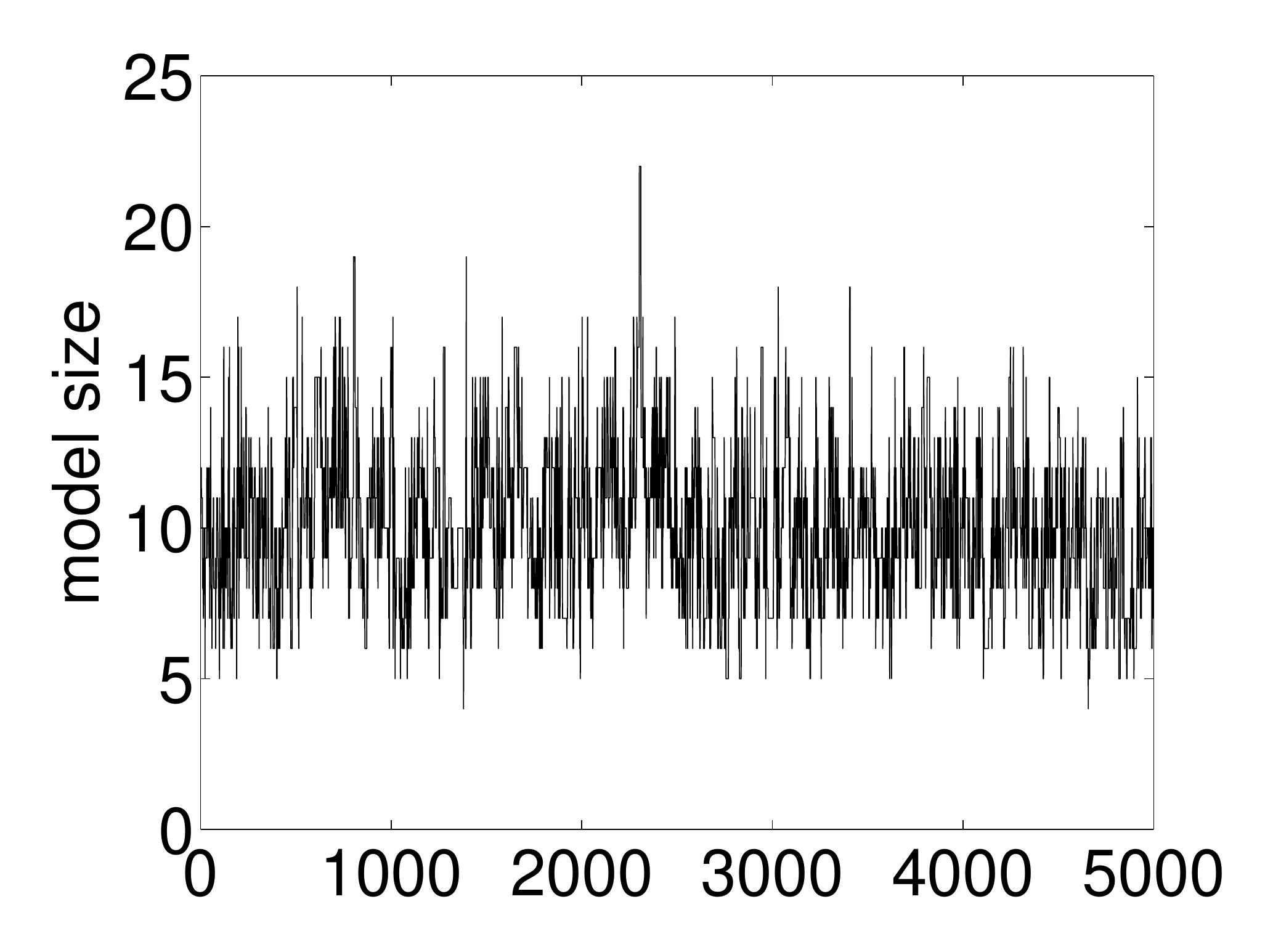} &
\includegraphics[trim=0mm 60mm 0mm 80mm, scale=0.25, clip]{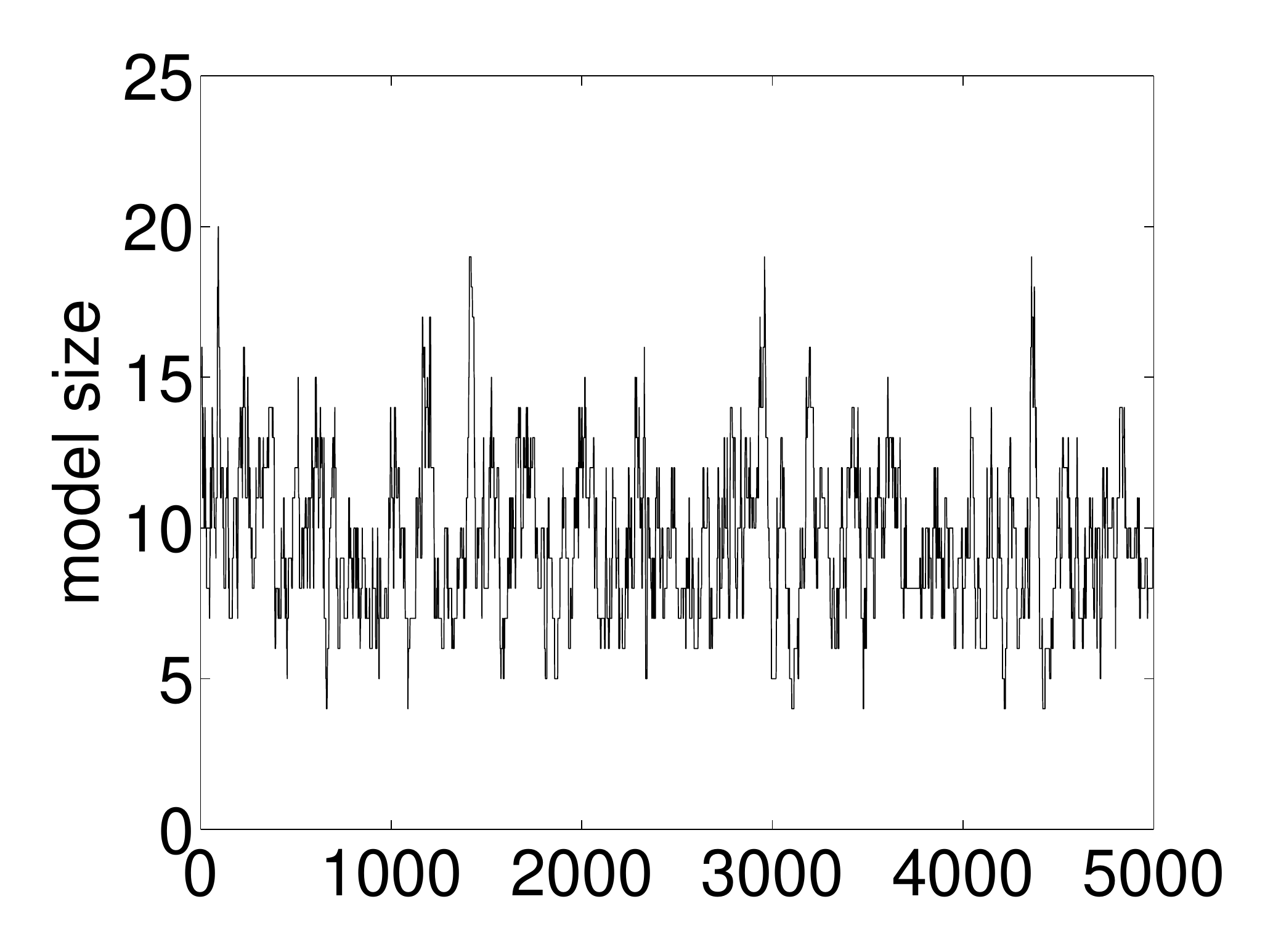} &
\includegraphics[trim=0mm 60mm 0mm 80mm, scale=0.25, clip]{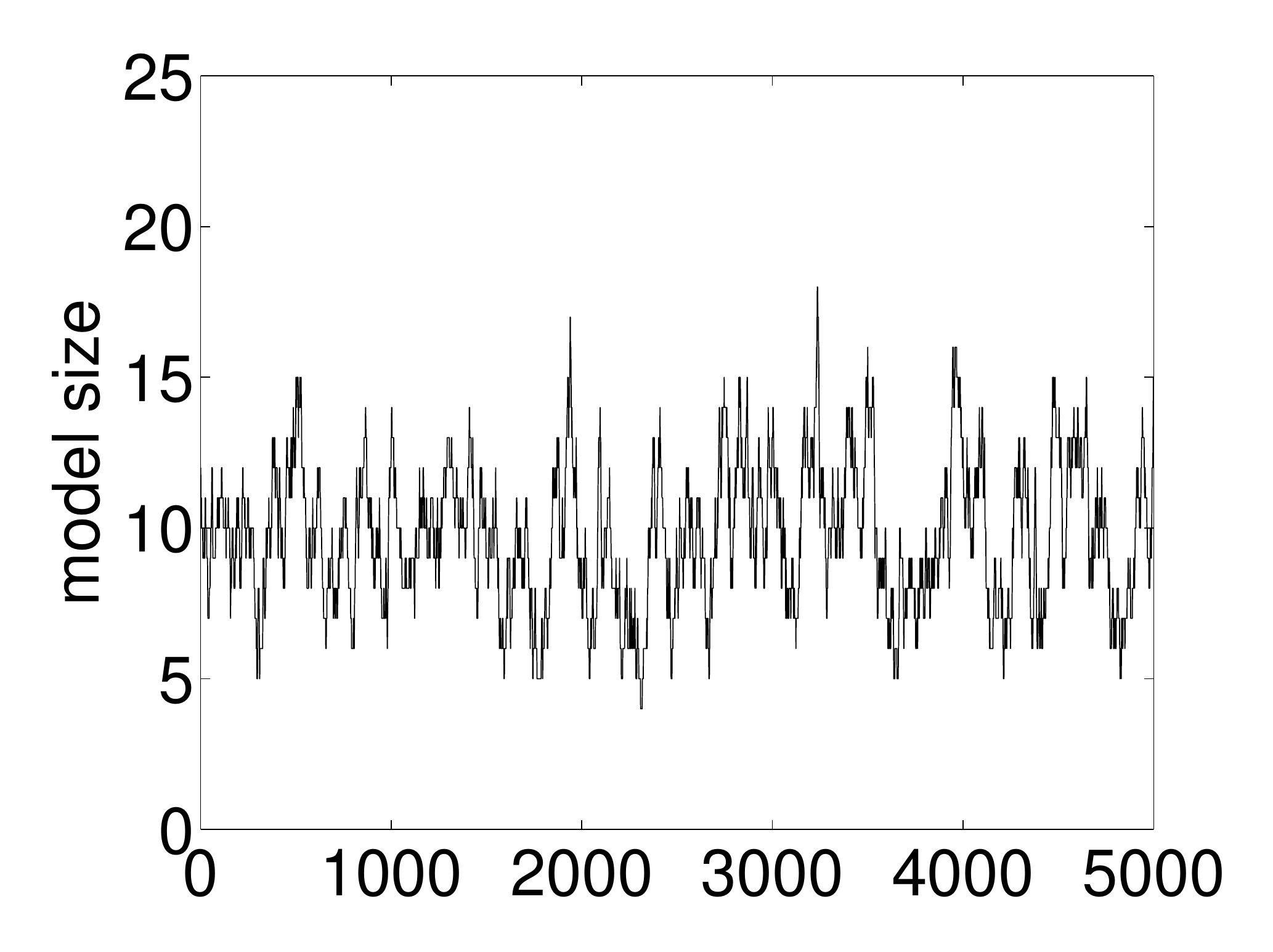}
\end{tabular}
\end{center}
\caption{\small Tecator data: model size for the last 5000 iterations from a single run of the IA-RAPA algorithm with  $\tau=0.45$ and two competitors 
}
\label{fig:tecator1_modelsize}
\end{figure}
As a comparison, 10 independent runs of a multi-move Metropolis-Hastings algorithm  with add, remove and swap moves
and the adaptive algorithm of  \cite{lamnisos12} were run.
 The multi-move sampler had an average ESS of 30\ 332 and the adaptive algorithm  had an average ESS of 40\ 000
(pretty much unaffected by the value of $\tau$ in the range (0.25,0.7)).
The best individual
adaptation algorithm had an ESS around 200\ 000 which represents roughly a six-fold increase over the multi-move sampler and roughly a five-fold increase over the adaptive algorithm.
The mixing of different algorithms with the tecator data is further illustrated in Figure~\ref{fig:tecator1_modelsize} which shows trace plots of the model size for a randomly chosen run. It is clear that the IA-RAPA algorithm leads to much better mixing than the two competitors.

\begin{figure}[h!]
\begin{center}
\begin{tabular}{ccc}
$\tau=0.35$ & $\tau=0.45$ & $\tau=0.55$\\
\includegraphics[trim=10mm 60mm 10mm 80mm, scale=0.25, clip]{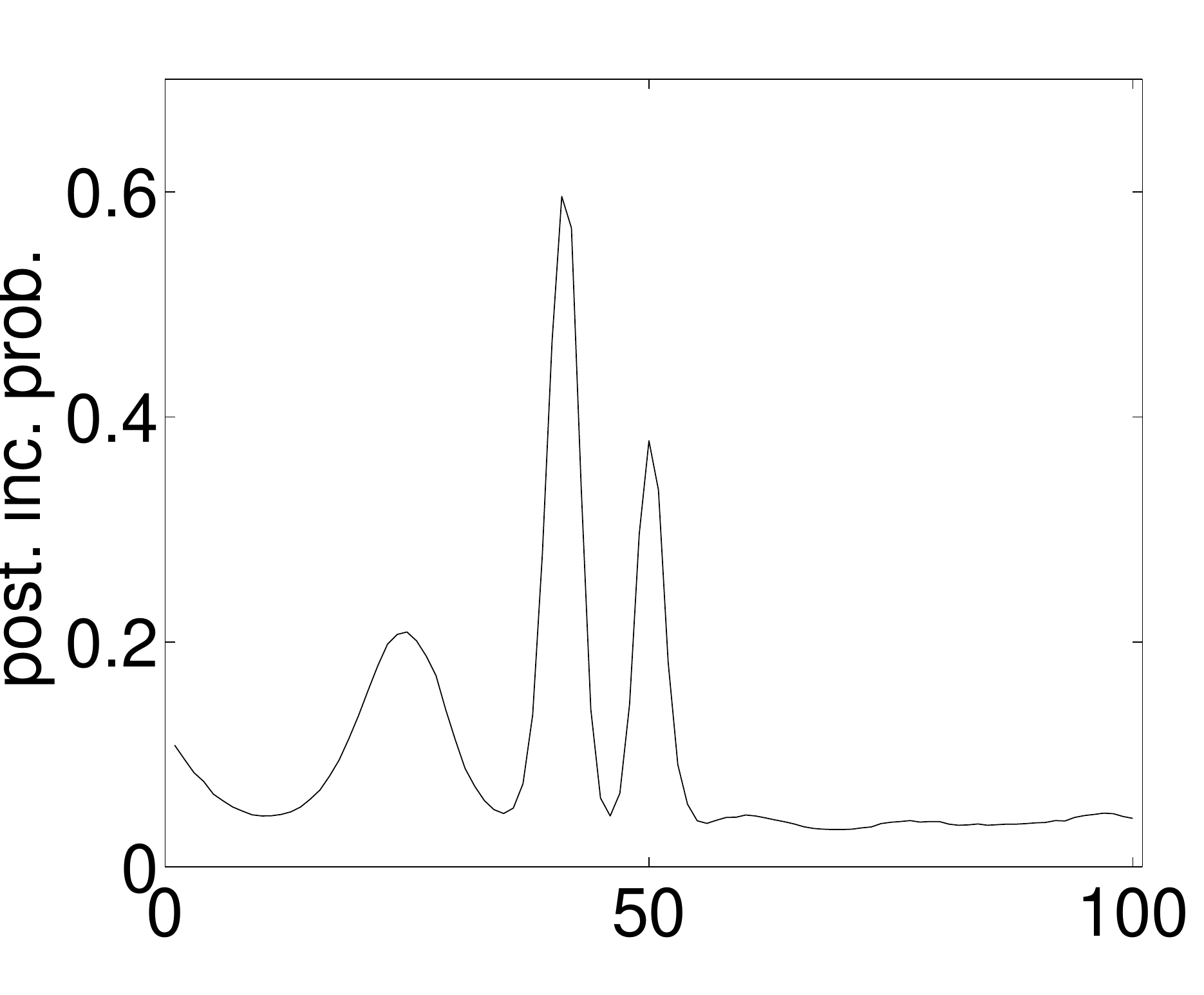} &
\includegraphics[trim=10mm 60mm 10mm 80mm, scale=0.25, clip]{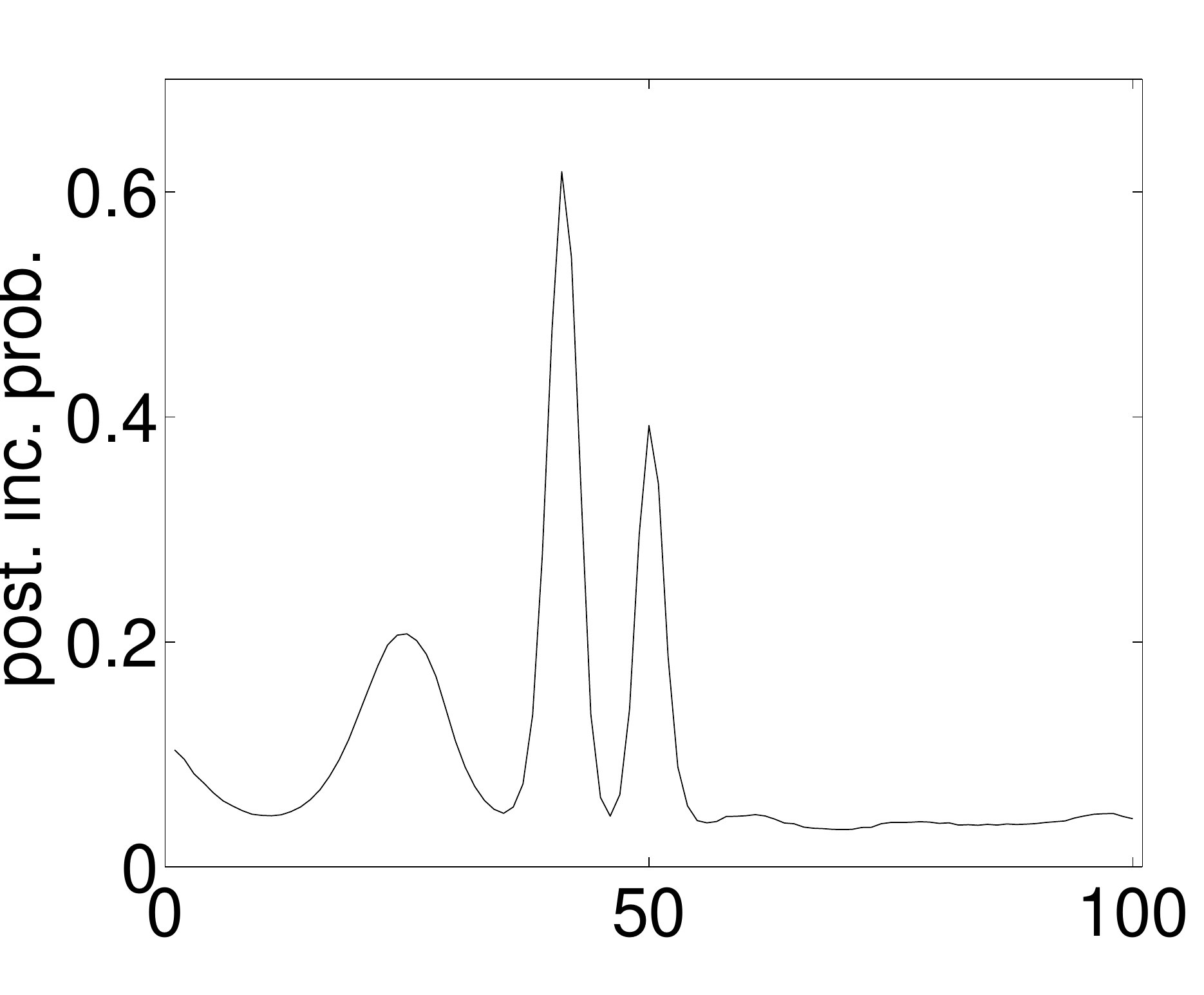} &
\includegraphics[trim=10mm 60mm 10mm 80mm, scale=0.25, clip]{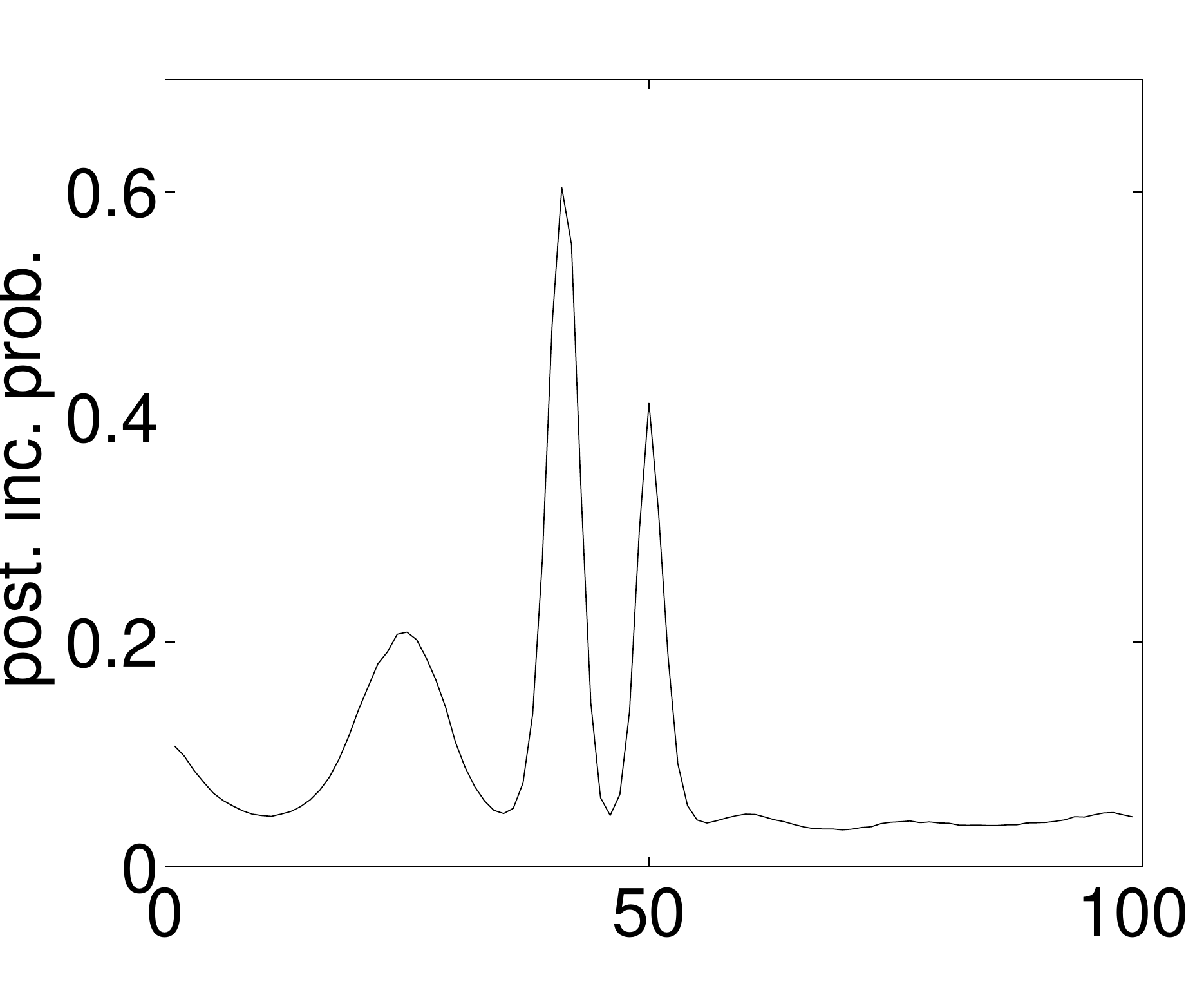}
\end{tabular}
\end{center}
\caption{\small Tecator data: PIP's estimated from a single run of the IA-RAPA algorithm with $\tau=0.35$, $\tau=0.45$ and $\tau=0.55$
}
\label{fig:tecator1_inc}
\end{figure}
\begin{figure}[h!]
\begin{center}
\begin{tabular}{ccc}
\includegraphics[trim=0mm 60mm 10mm 80mm, scale=0.25, clip]{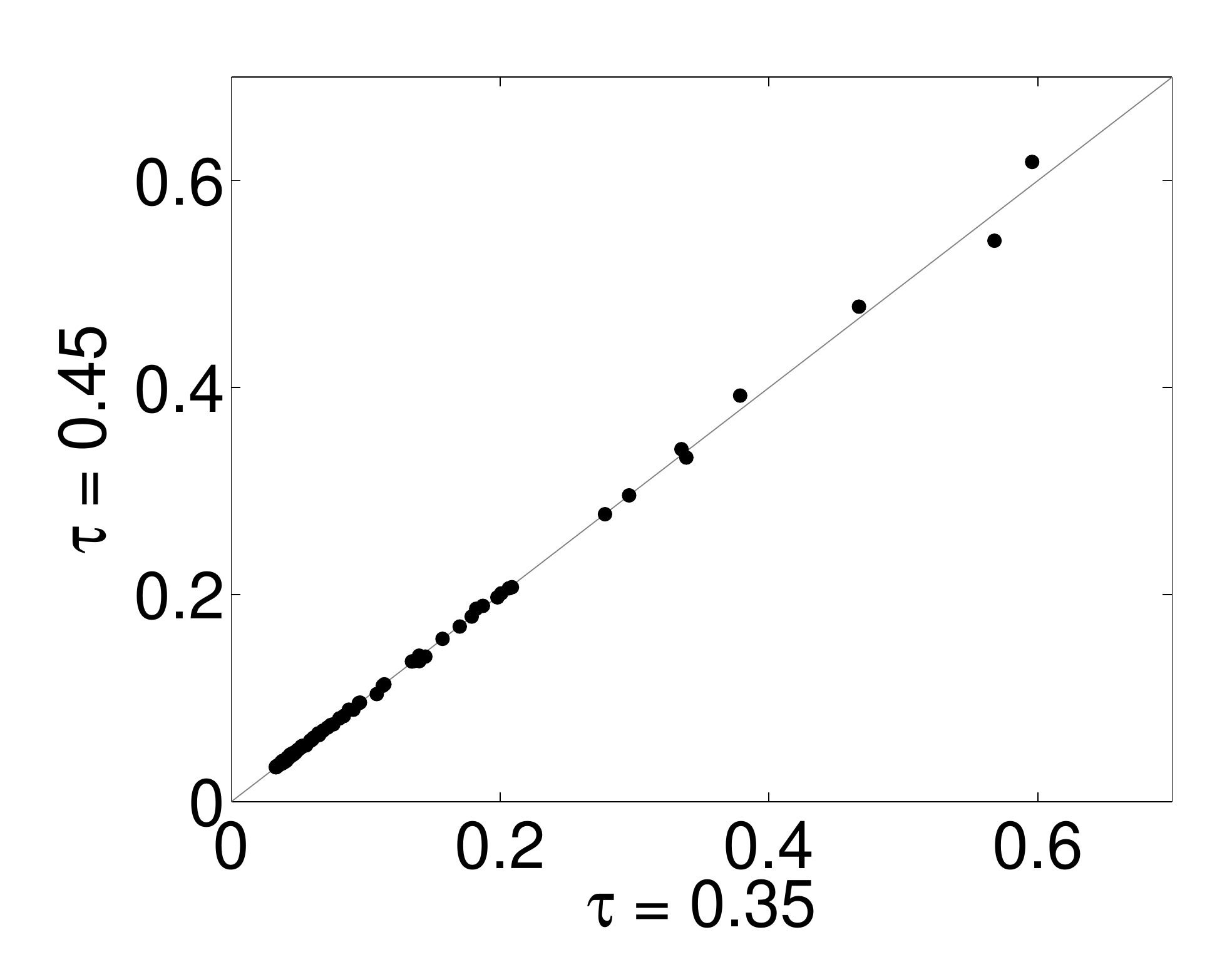} &
\includegraphics[trim=0mm 60mm 10mm 80mm, scale=0.25, clip]{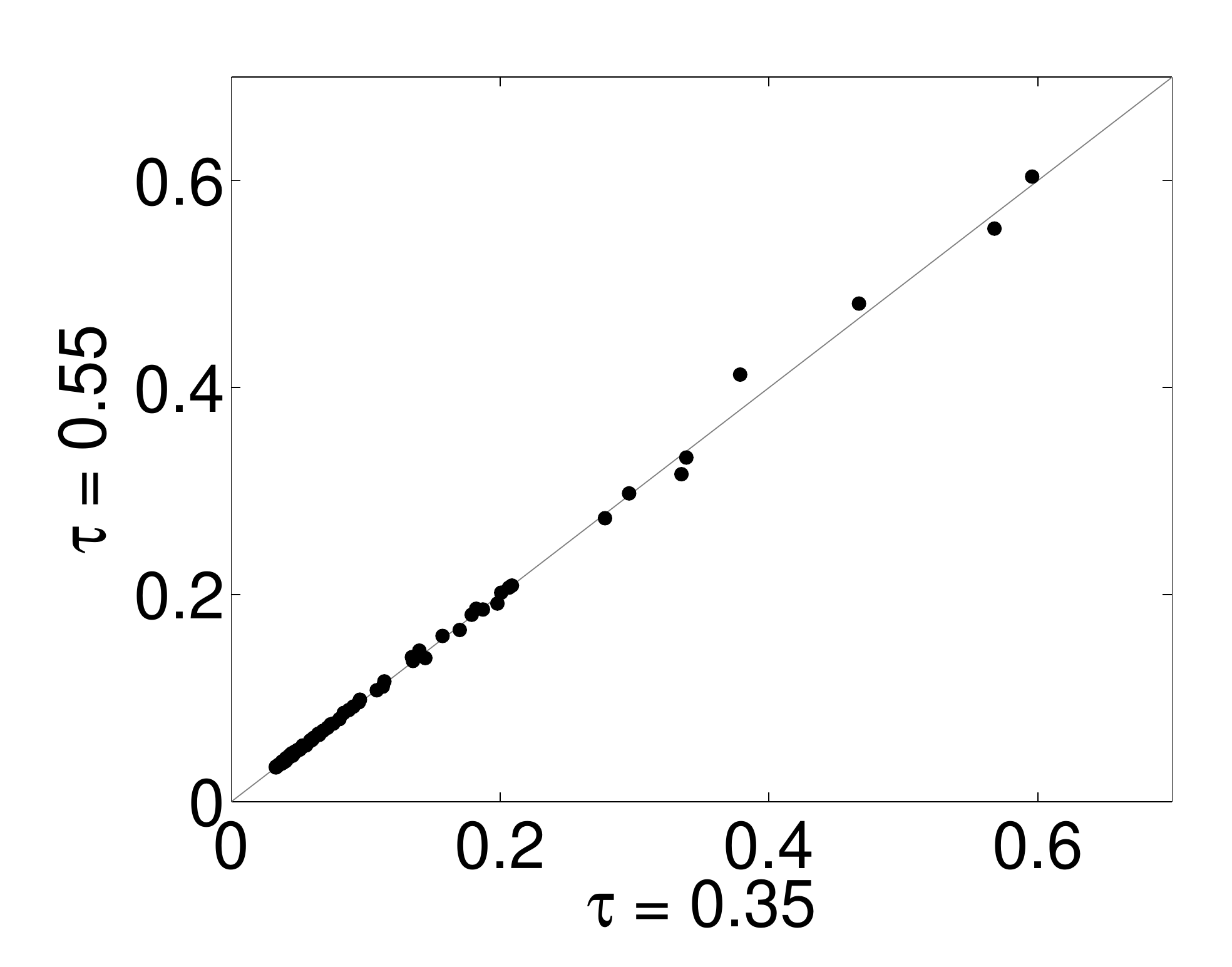} &
\includegraphics[trim=0mm 60mm 10mm 80mm, scale=0.25, clip]{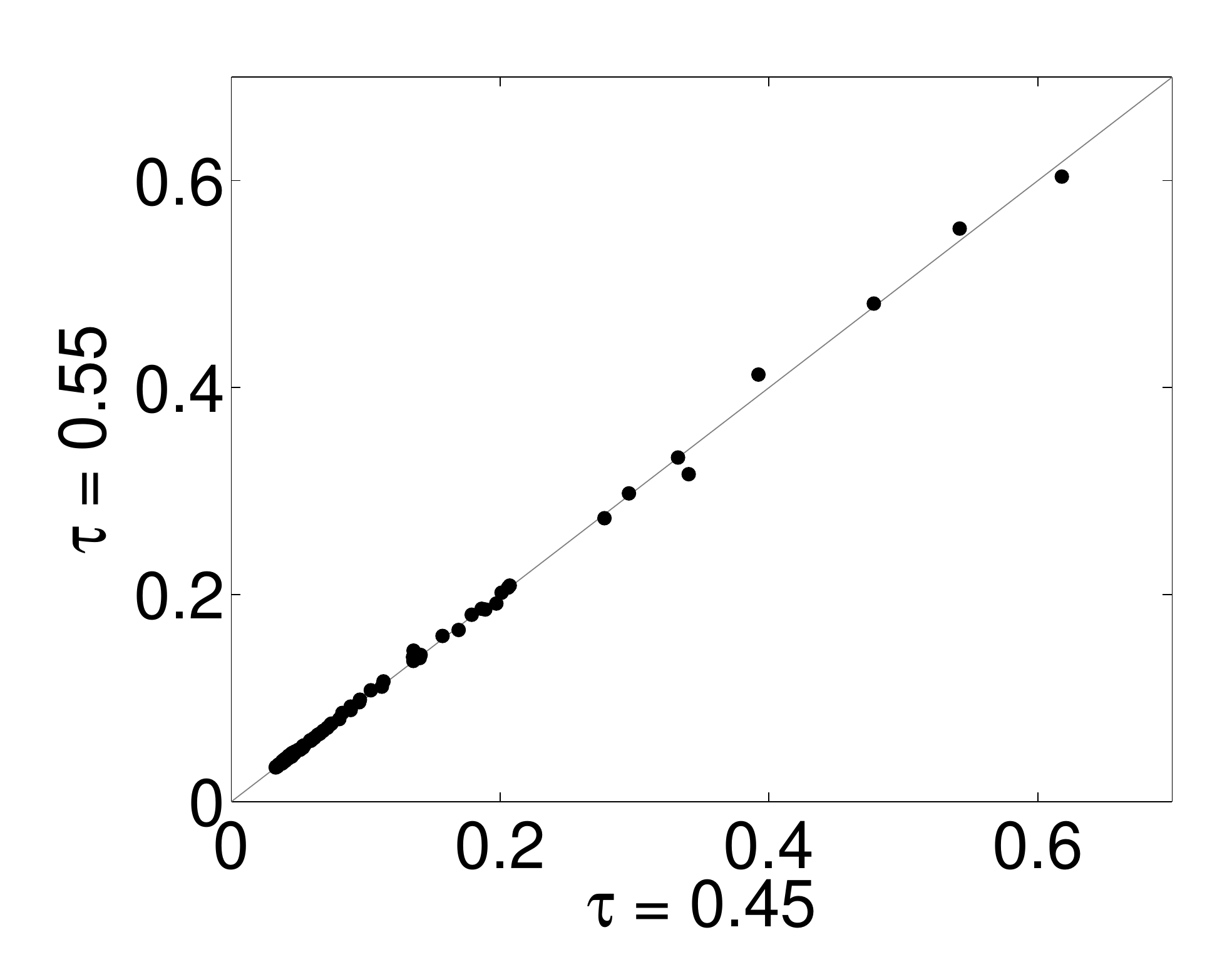}
\end{tabular}
\end{center}
\caption{\small Tecator data: scatter plots of pairs of PIP's estimated from a single run of the IA-RAPA algorithm with $\tau=0.35$, $\tau=0.45$ and $\tau=0.55$. The thin solid line is $y=x$
}
\label{fig:tecator1_inc_cross}
\end{figure}
Insight into the behaviour of the algorithm is provided by looking at the results of
single runs of the IA-RAPA algorithm with different values of $\tau$ with
a burn-in period of 100\ 000 iterations, a subsequent sample of
1 million iterations taken and no thinning.
Figure~\ref{fig:tecator1_inc} shows the PIP's  and Figure~\ref{fig:tecator1_inc_cross} shows scatter-plots of pairs of the estimated posterior inclusion probabilities with the different values of $\tau$.  These indicate very strong agreement across the runs of the individual adaptation algorithms with different $\tau$.

\begin{figure}[h!]
\begin{center}
\begin{tabular}{ccc}
$\tau=0.35$ & $\tau=0.45$ & $\tau=0.55$\\
\includegraphics[trim=10mm 60mm 10mm 80mm, scale=0.25, clip]{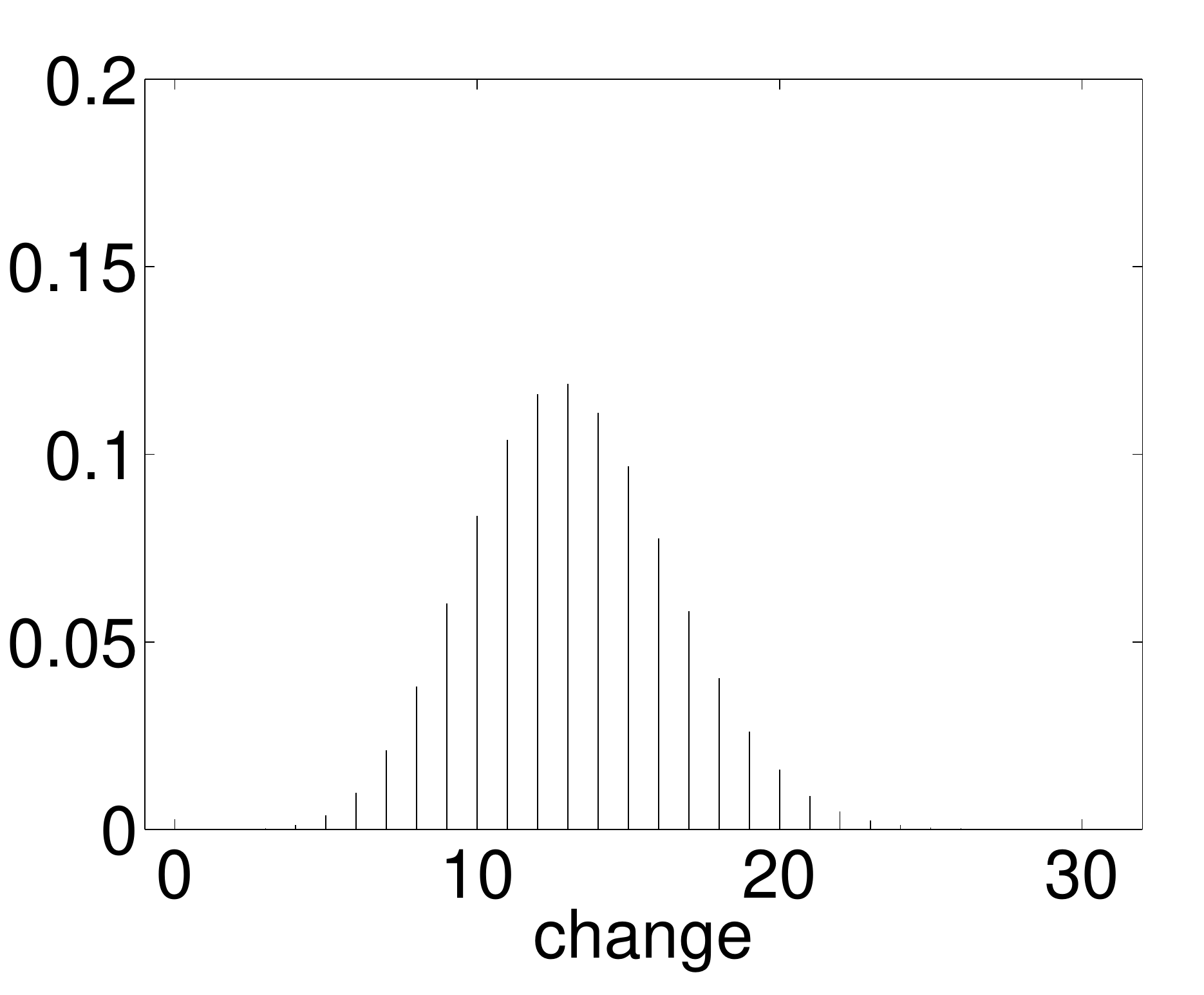} &
\includegraphics[trim=10mm 60mm 10mm 80mm, scale=0.25, clip]{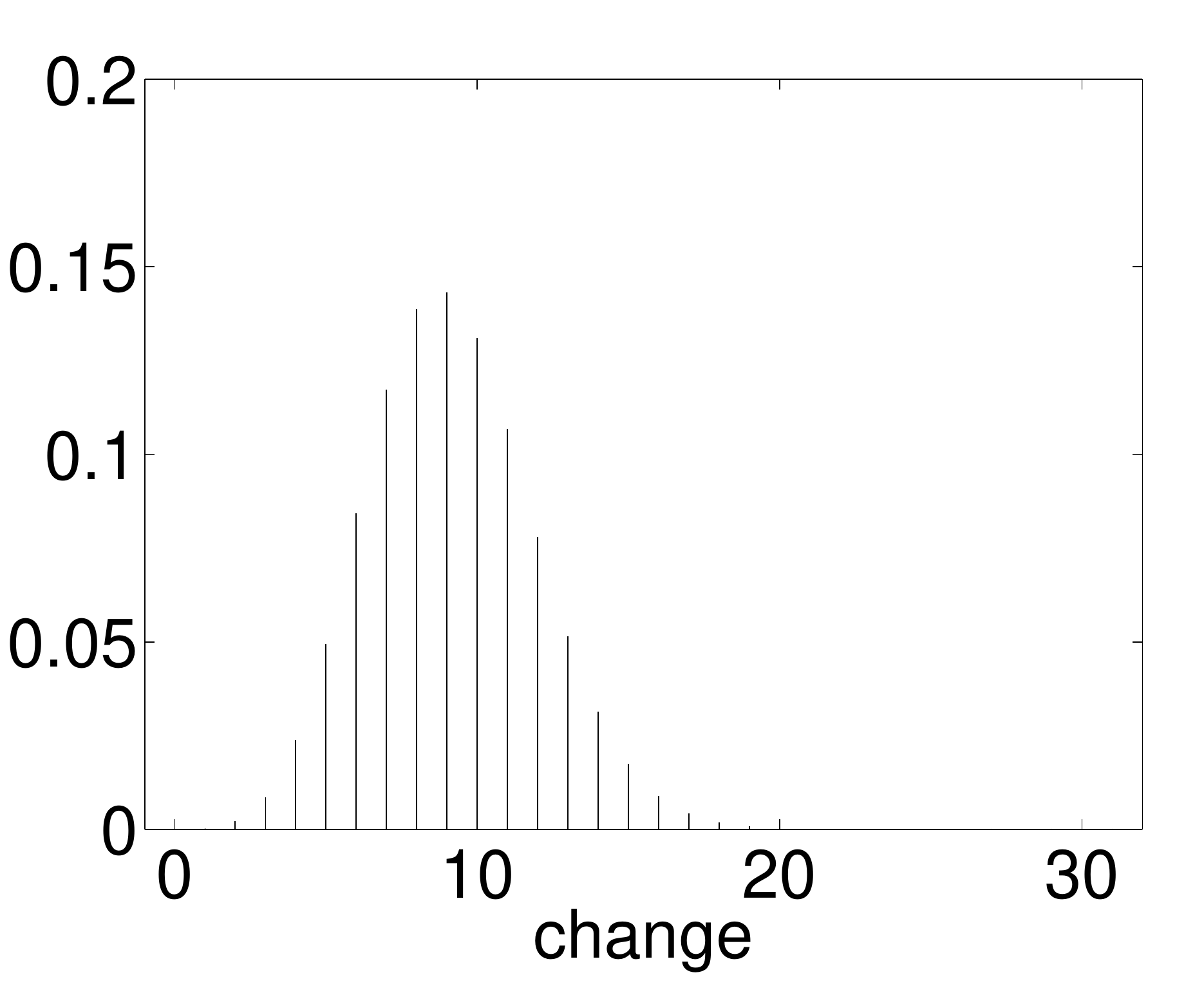} &
\includegraphics[trim=10mm 60mm 10mm 80mm, scale=0.25, clip]{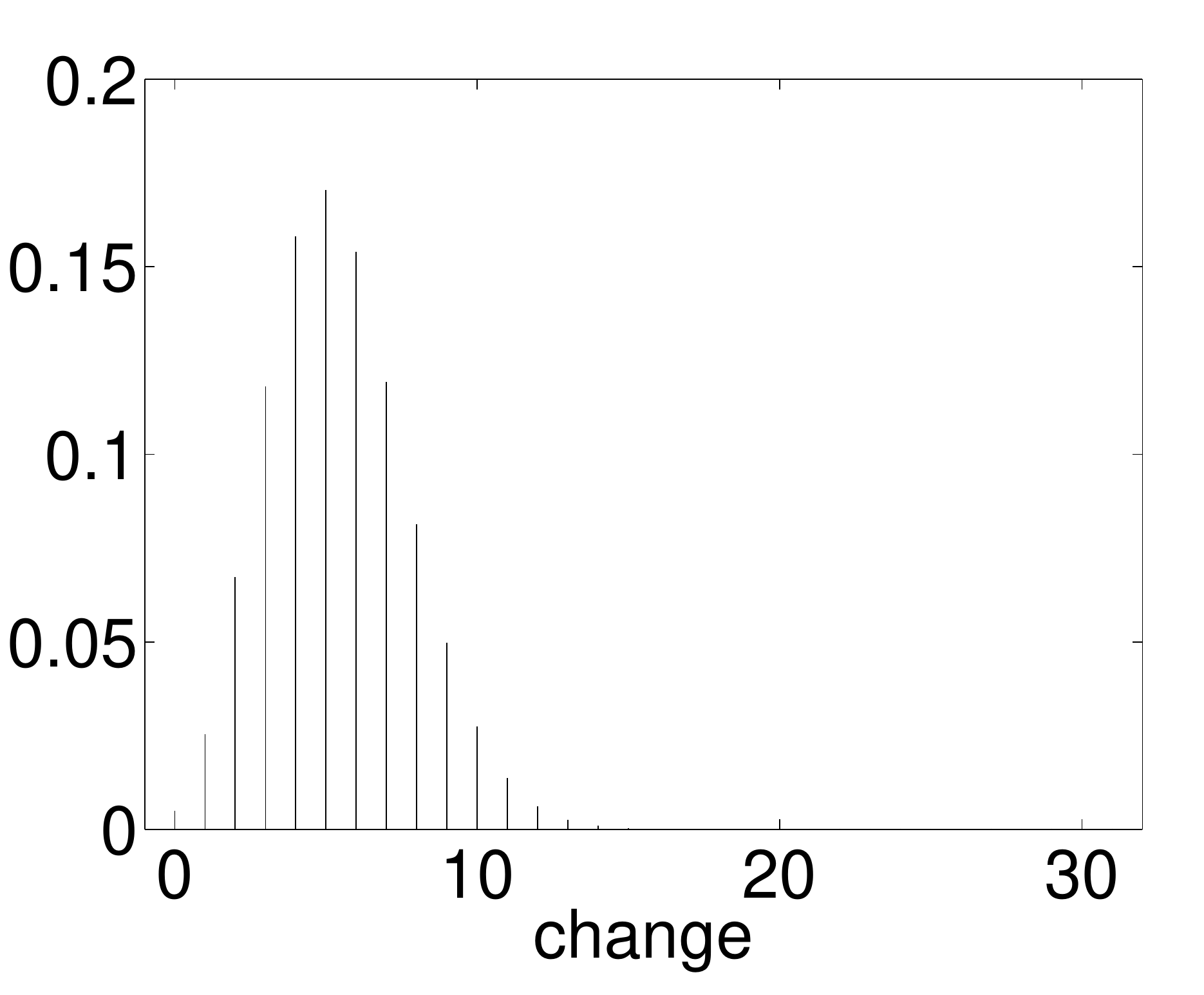}
\end{tabular}
\end{center}
\caption{\small Tecator data: empirical probability mass function of the number of variables proposed to be changed at each iteration during a single run of the IA-RAPA algorithm with $\tau=0.35$, $\tau=0.45$ and $\tau=0.55$
}
\label{fig:tecator1_change}
\end{figure}
The empirical probability mass function of the number of variables proposed to be changed at each step of the algorithm is shown in Figure~\ref{fig:tecator1_change}. The modal value is 14 for $\tau=0.35$ with a sizeable spread of values from 5 to 22. This illustrates that relatively large changes in the model are possible in this example. The location and spread of the distribution becomes smaller as $\tau$ increases and less ambitious moves are proposed.

\begin{figure}[h!]
\begin{center}
\begin{tabular}{ccc}
 $\tau=0.35$ & $\tau=0.45$ & $\tau=0.55$\\
 \includegraphics[trim=0mm 60mm 10mm 80mm, scale=0.25, clip]{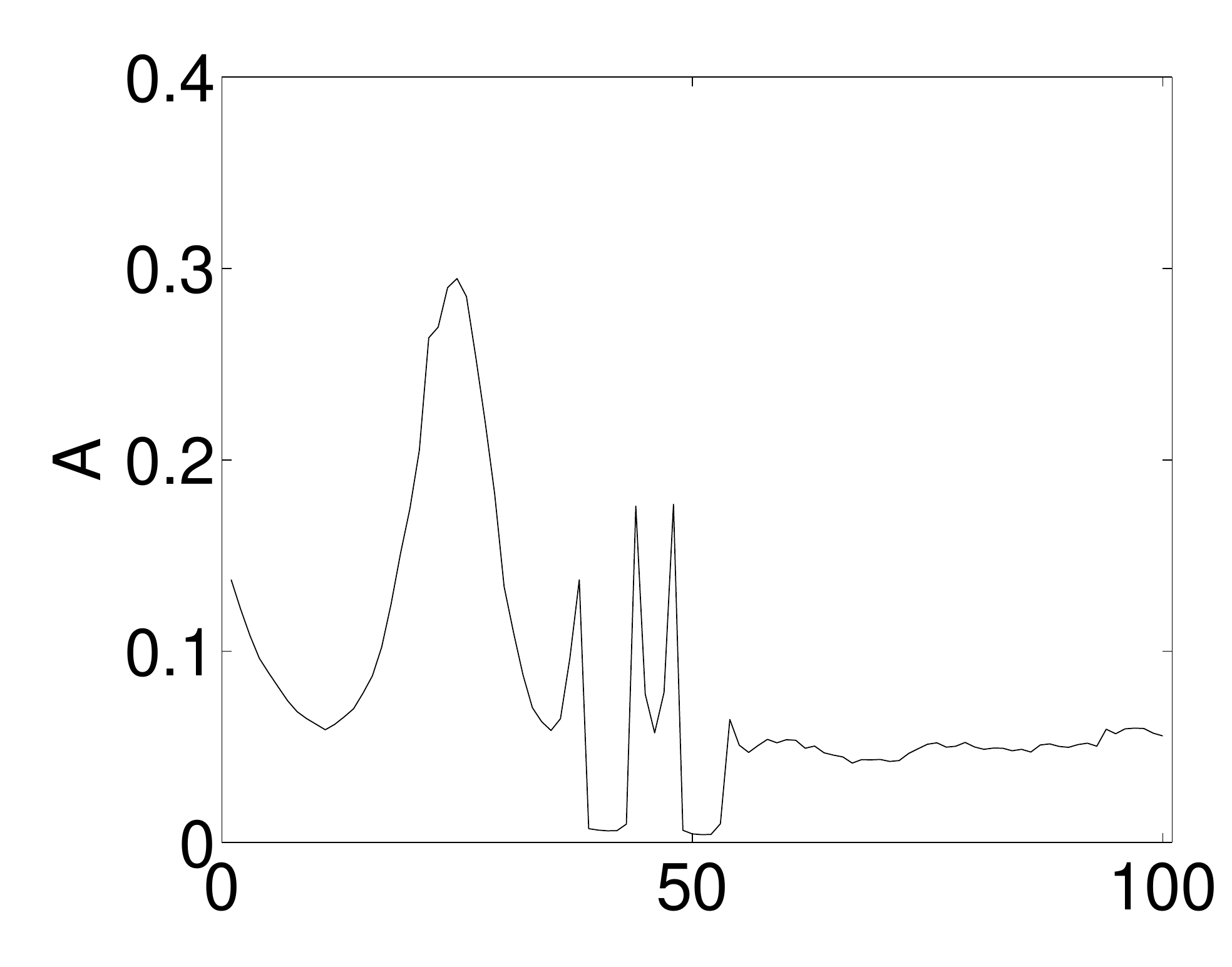} &
\includegraphics[trim=0mm 60mm 10mm 80mm, scale=0.25, clip]{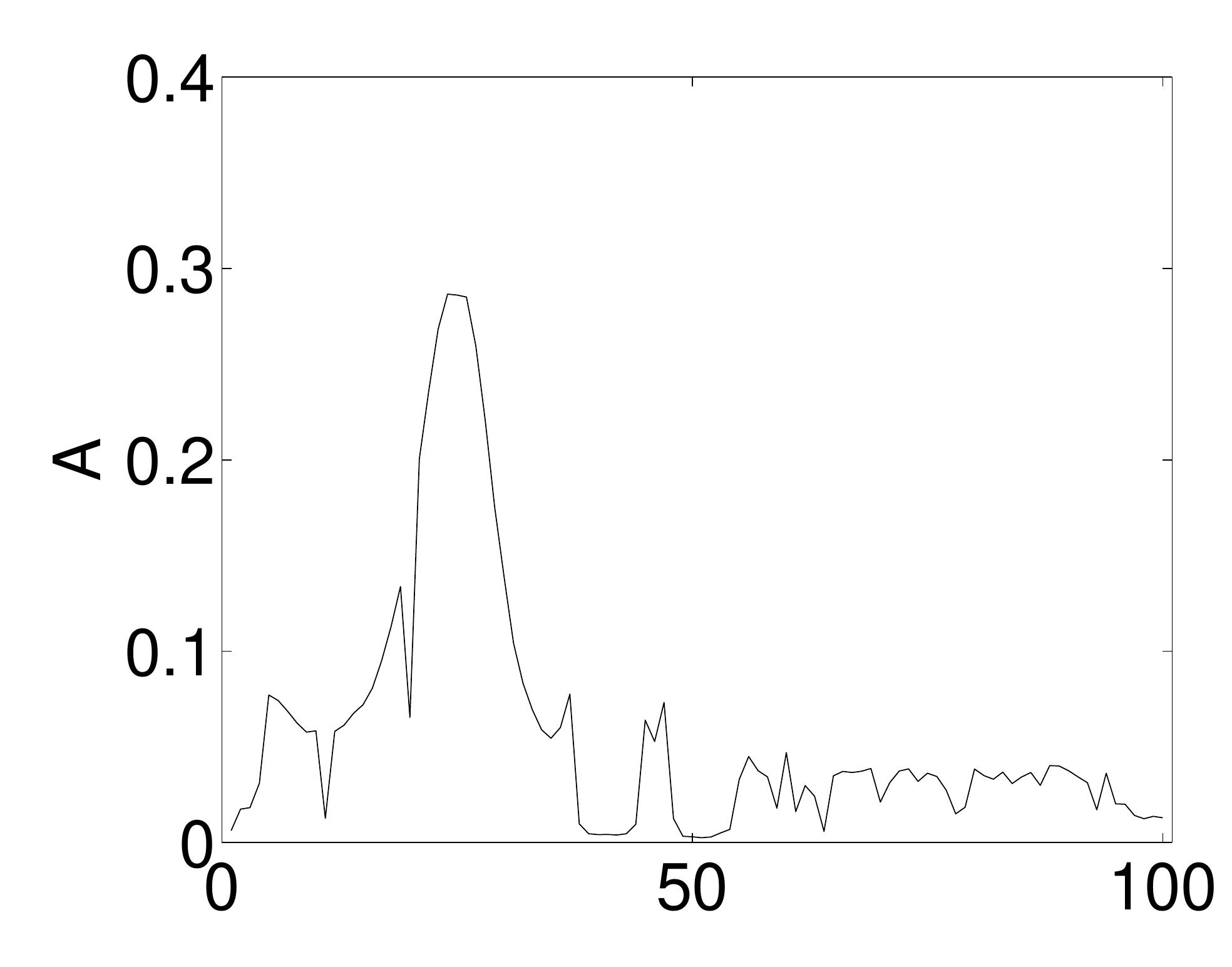} &
\includegraphics[trim=0mm 60mm 10mm 80mm, scale=0.25, clip]{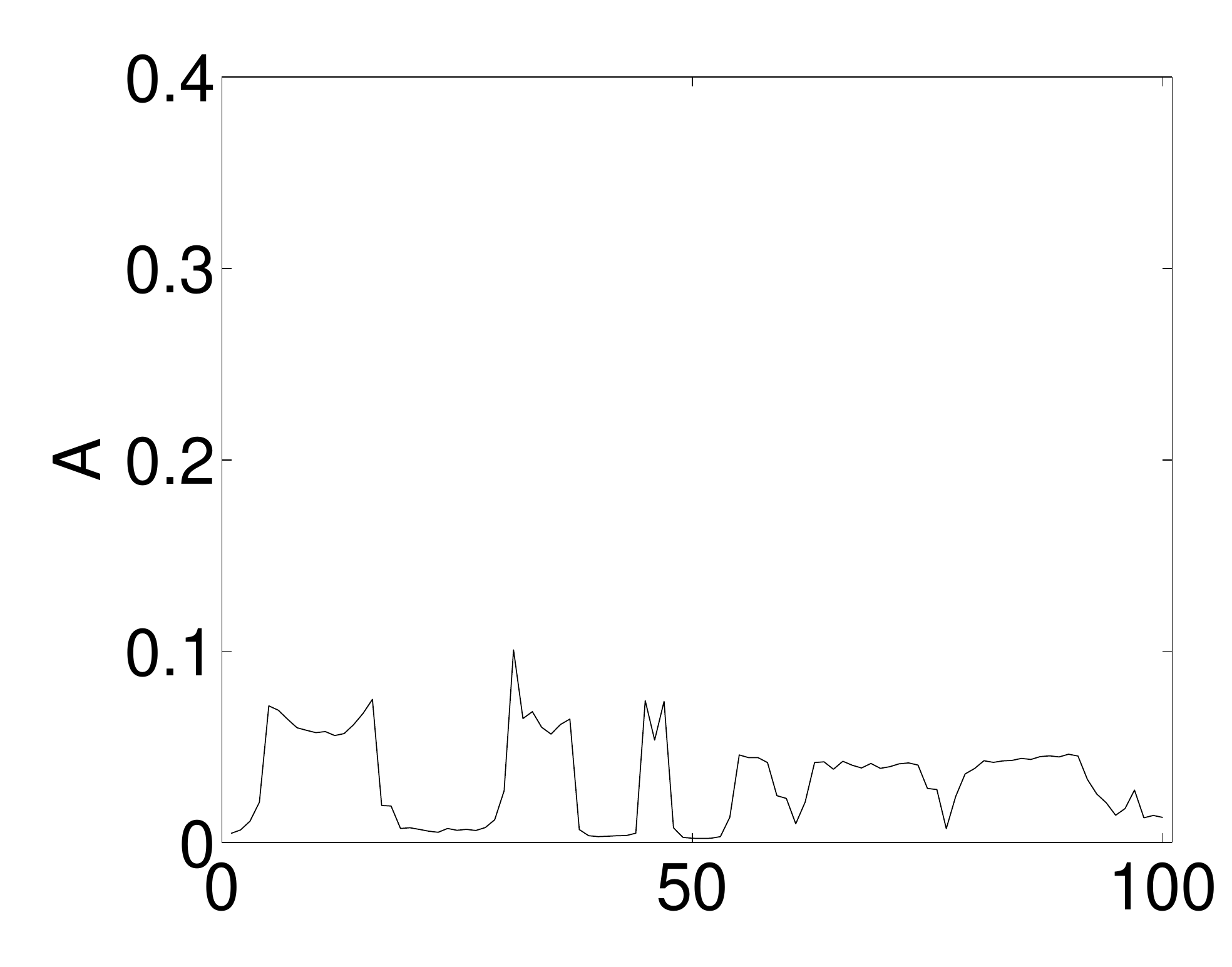}\\
\includegraphics[trim=0mm 60mm 10mm 80mm, scale=0.25, clip]{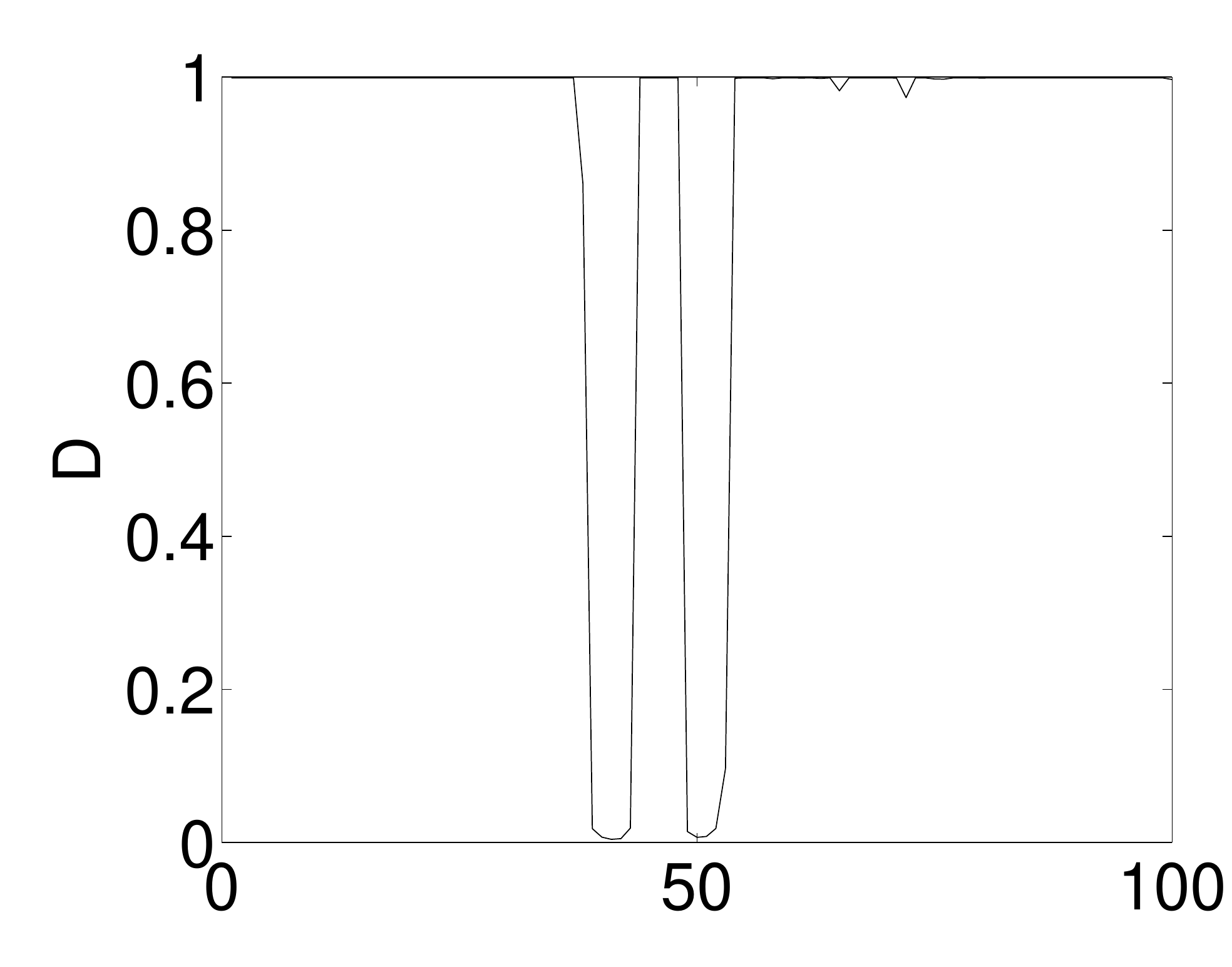} &
\includegraphics[trim=0mm 60mm 10mm 80mm, scale=0.25, clip]{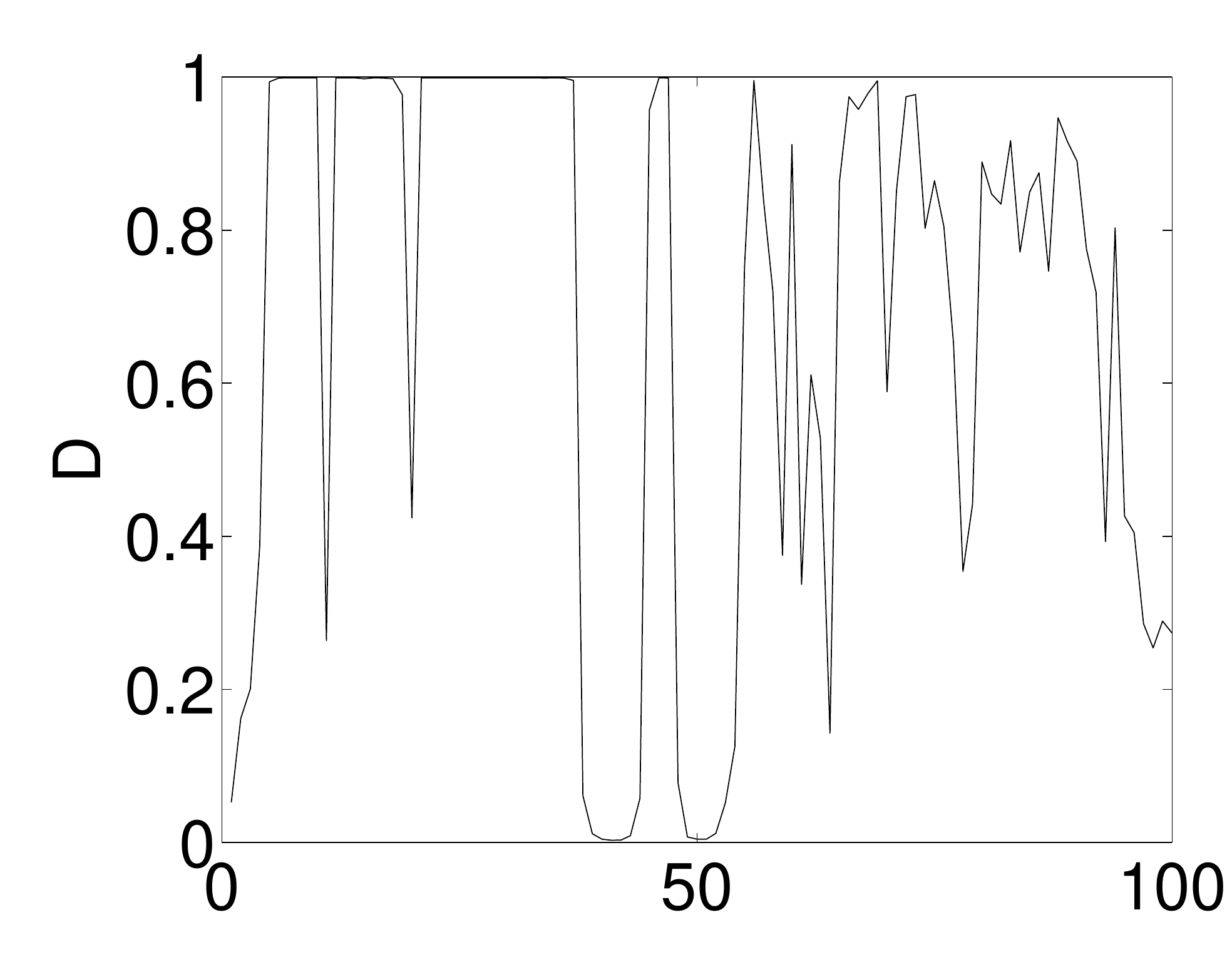} &
\includegraphics[trim=0mm 60mm 10mm 80mm, scale=0.25, clip]{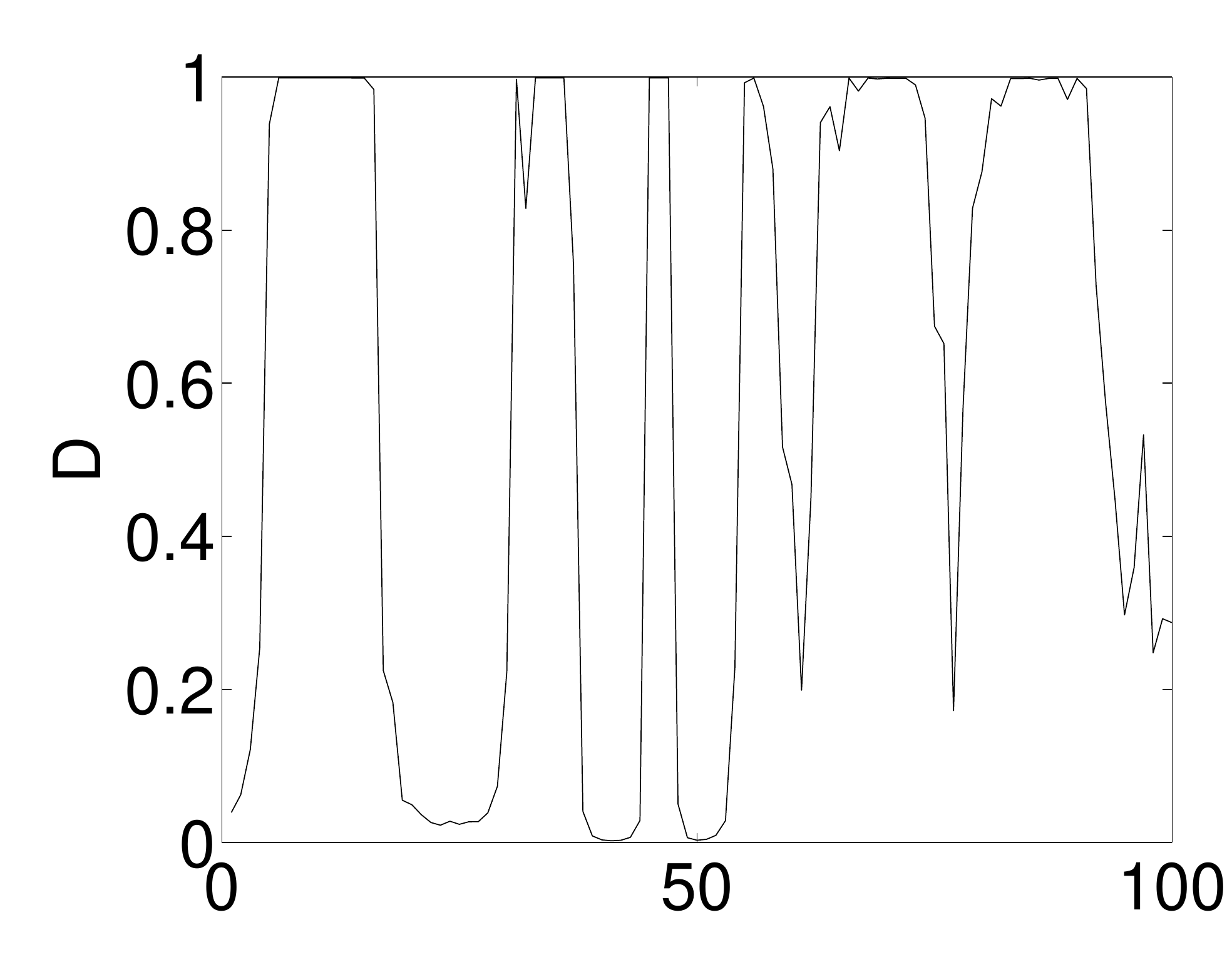}
\end{tabular}
\end{center}
\caption{\small Tecator data: values of $A_j$ and $D_j$ at the end of a single run of the IA-RAPA algorithm with $\tau=0.35$, $\tau=0.45$ and $\tau=0.55$
}
\label{fig:tecator1_zetaAD}
\end{figure}

The values of $A$ and $D$ for a single run of the algorithm with different values of $\tau$ are shown in
Figures~\ref{fig:tecator1_zetaAD}. Overall, the values of $A$ tend to decrease as $\tau$ increases.
Therefore, the algorithm proposes less ambitious moves which leads to a larger
average mutation rate. The values of
$D_j$ tend to be close to 0 or 1 when $\tau=0.35$. The value is close to zero for variables which have a higher inclusion probability whereas $D_j$ is close to one for variables which have a lower PIP. Therefore, the algorithm will usually propose to remove variables with low PIP's if they are currently included in the model and tend to not propose removing variables with high PIP's. This type of behaviour is critical for rapid mixing in this type of problem. If a variable has a low PIP, say 0.05 or 0.1, the best mixing would occur if this variable was removed from the model as quickly as possible after being added (whilst maintaining the correct PIP).
 The values of $D$ become less extreme as $\tau$ increases.

The values of $A_j$ and $D_j$ at the end of each run tend to be different (although, many final values of $A_j$ and $D_j$ will be similar across different runs). As we have already mentioned, the convergence of the sampler does not depend on the convergence of the $A_j$'s or $D_j$'s. However, the ratio $A_j/D_j$ tends to have a consistent value across different runs and different values of $\tau$.
\begin{figure}[h!]
\begin{center}
\begin{tabular}{ccc}
$\tau=0.35$ & $\tau=0.45$ & $\tau=0.55$\\
\includegraphics[trim=5mm 60mm 10mm 80mm, scale=0.25, clip]{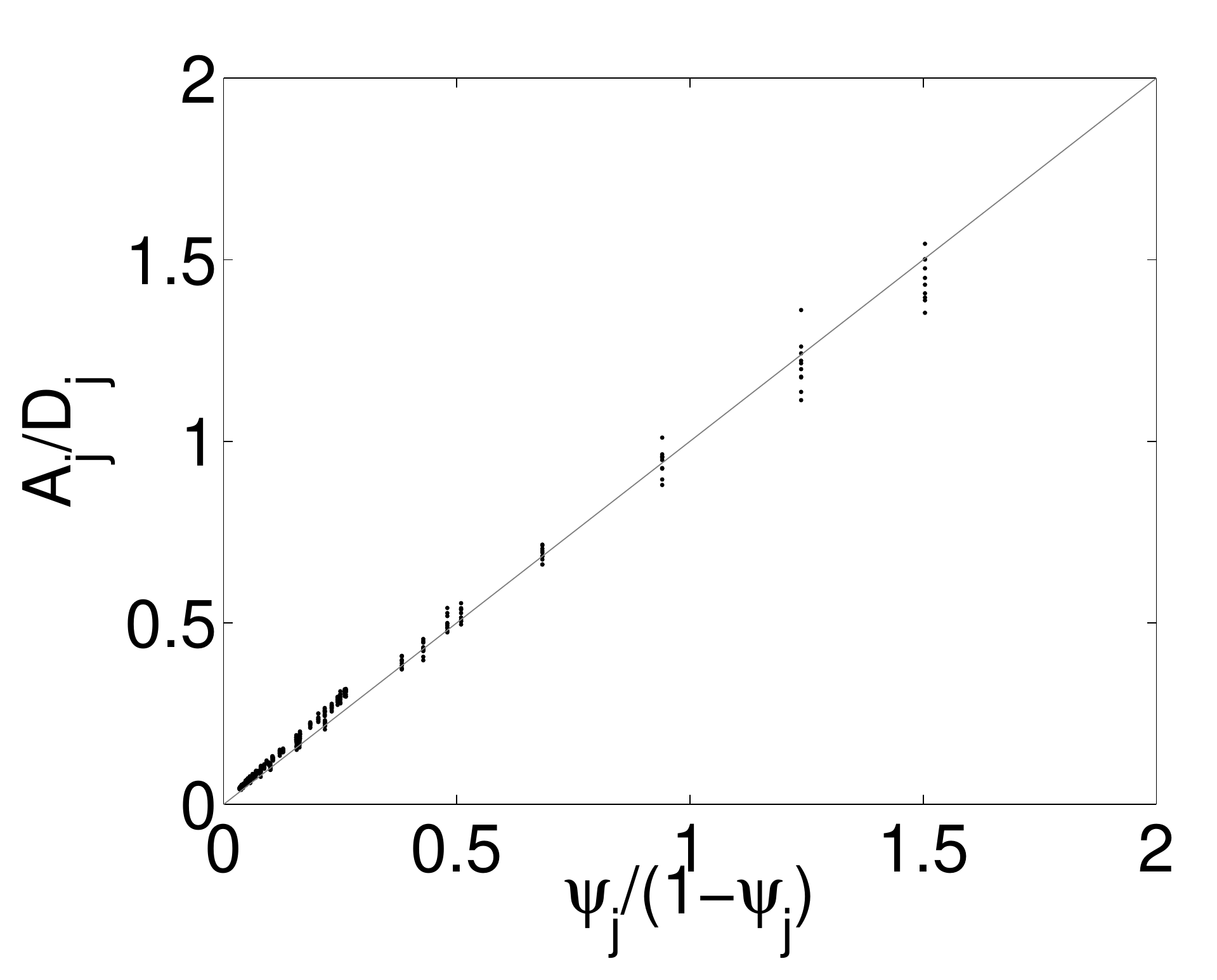} &
\includegraphics[trim=5mm 60mm 10mm 80mm, scale=0.25, clip]{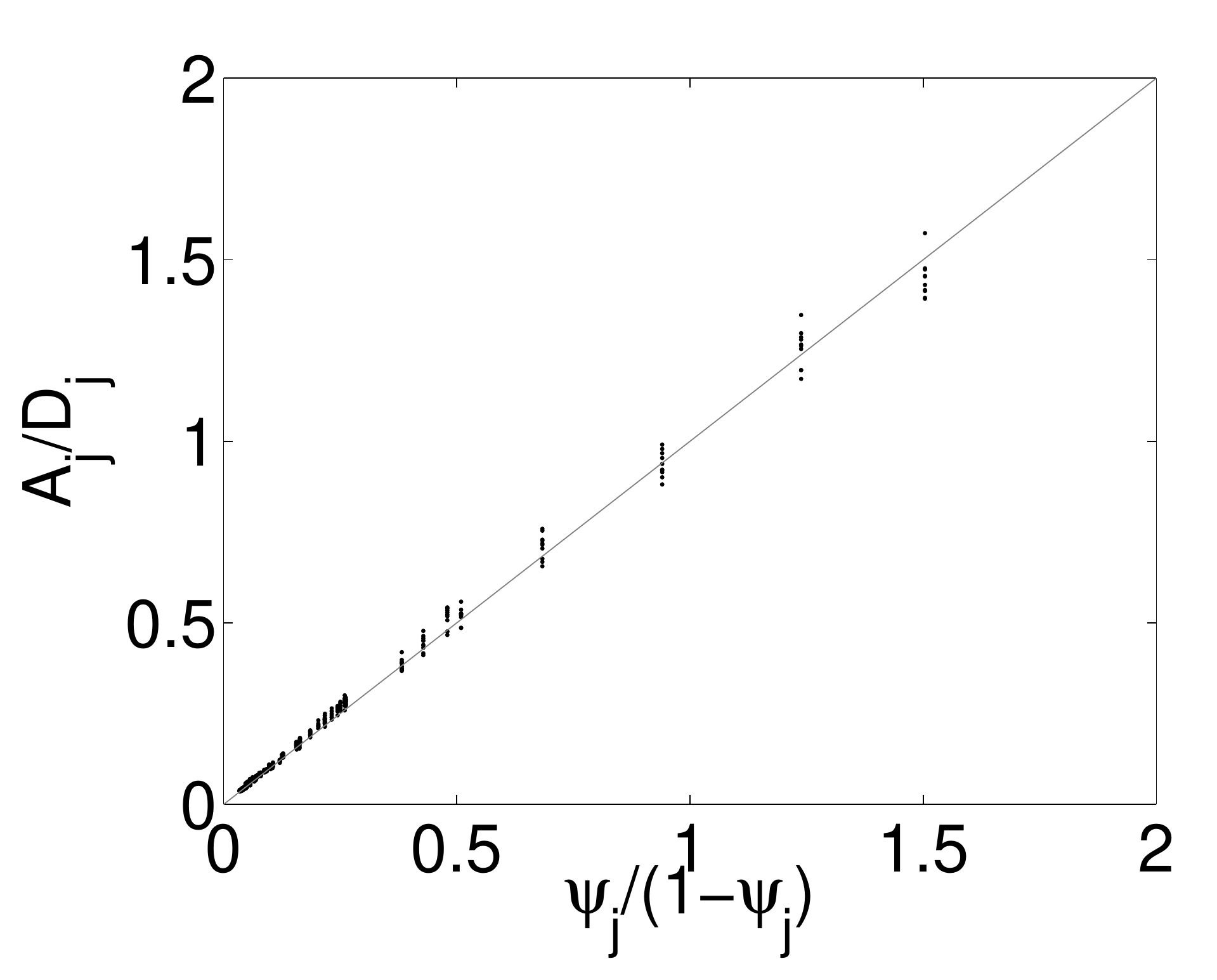} &
\includegraphics[trim=5mm 60mm 10mm 80mm, scale=0.25, clip]{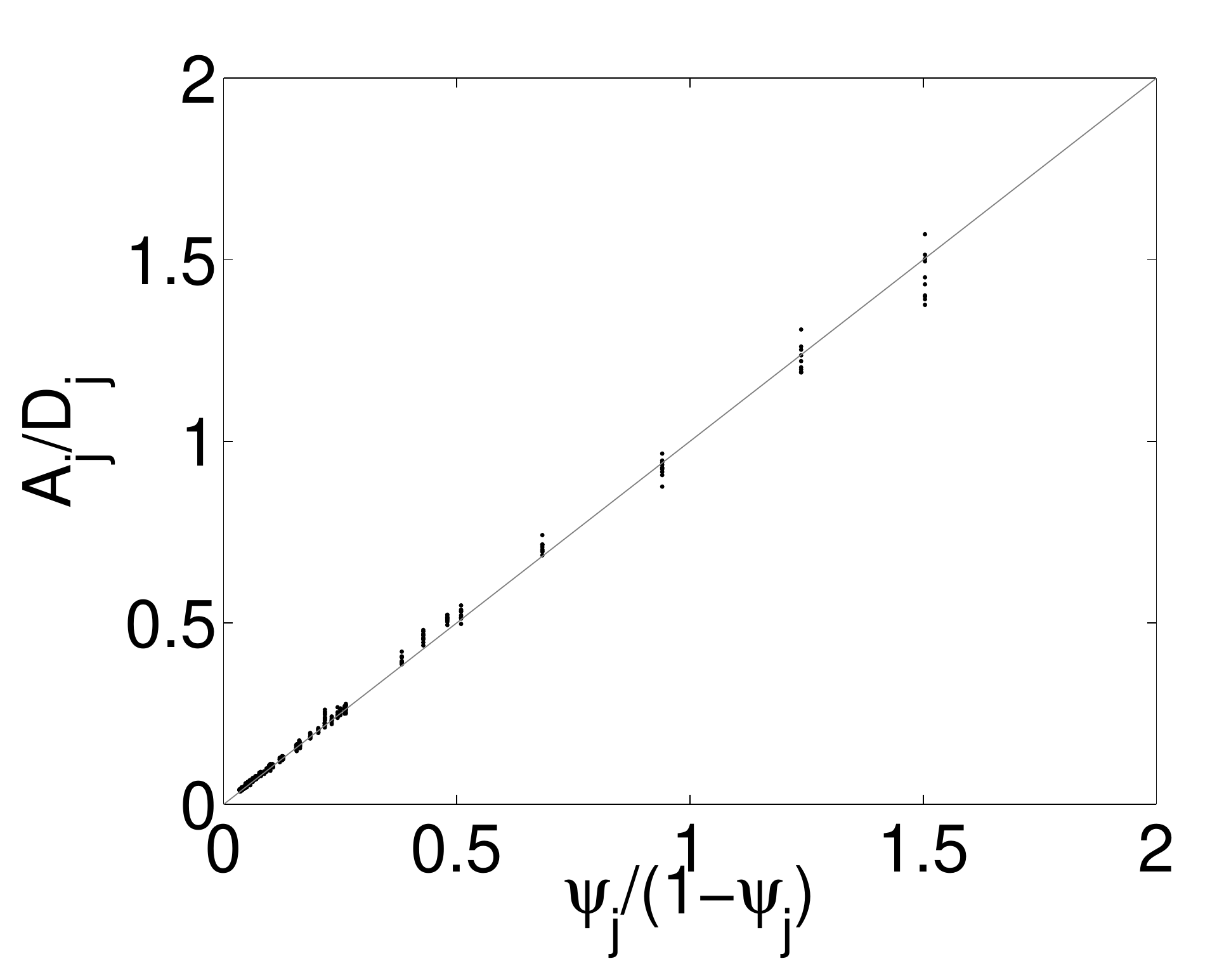}
\end{tabular}
\end{center}
\caption{\small Tecator data: scatterplot of $A_j/D_j$  at the end of 10 different runs of the
IA-RAPA algorithm
against $\psi_j/(1-\psi_j)$ where $\psi_j$ is the PIP of the $j$-th regressor calculated using all runs
with $\tau=0.35$, $\tau=0.45$ and $\tau=0.55$
}
\label{fig:tecator1_ratio2}
\end{figure}
Figure~\ref{fig:tecator1_ratio2} shows that $A_j/D_j$ is typically very close to $\psi_j/(1-\psi_j)$ where $\psi_j$ is the PIP of the $j$-th variable.
As a simple explanation of this effect, consider a posterior for $\gamma$ which is independent: then the Metropolis-Hastings acceptance rate of both adding and removing a variable will be 1 if $A_j/D_j=\psi_j/(1-\psi_j)$ and so this maximizes the overall acceptance rate. Of course, the posterior distribution will typically be far from independent and
this chain will not lead to optimal performance in general.

\subsection{PCR Data}

\cite{BoRe12} described a variable selection problem with 22\ 576 variables and 60 observations on two inbred mouse populations. The covariates are gender and gene expression  measurements for 22\ 575 genes. Using quantitative real-time polymerase chain reaction (PCR) several physiological phenotypes are recorded.
\begin{figure}[h!]
\begin{center}
5 chains\\
\includegraphics[scale=0.8, clip, trim=10mm 0mm 20mm 245mm ]{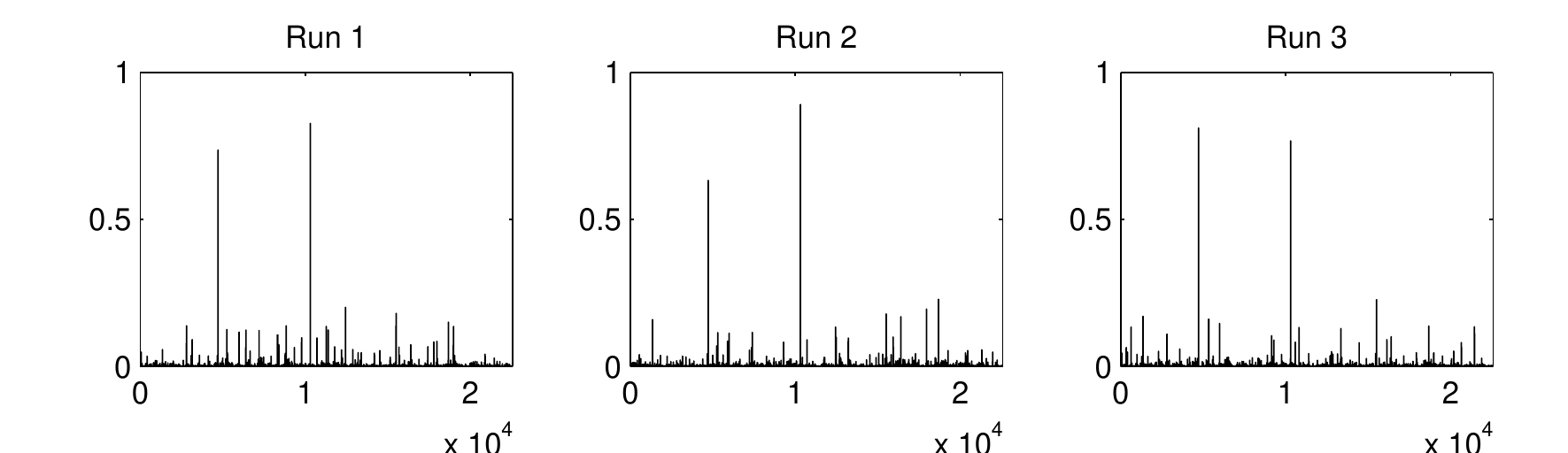}\\
25 chains\\
\includegraphics[scale=0.8, clip, trim=10mm 0mm 20mm 245mm ]{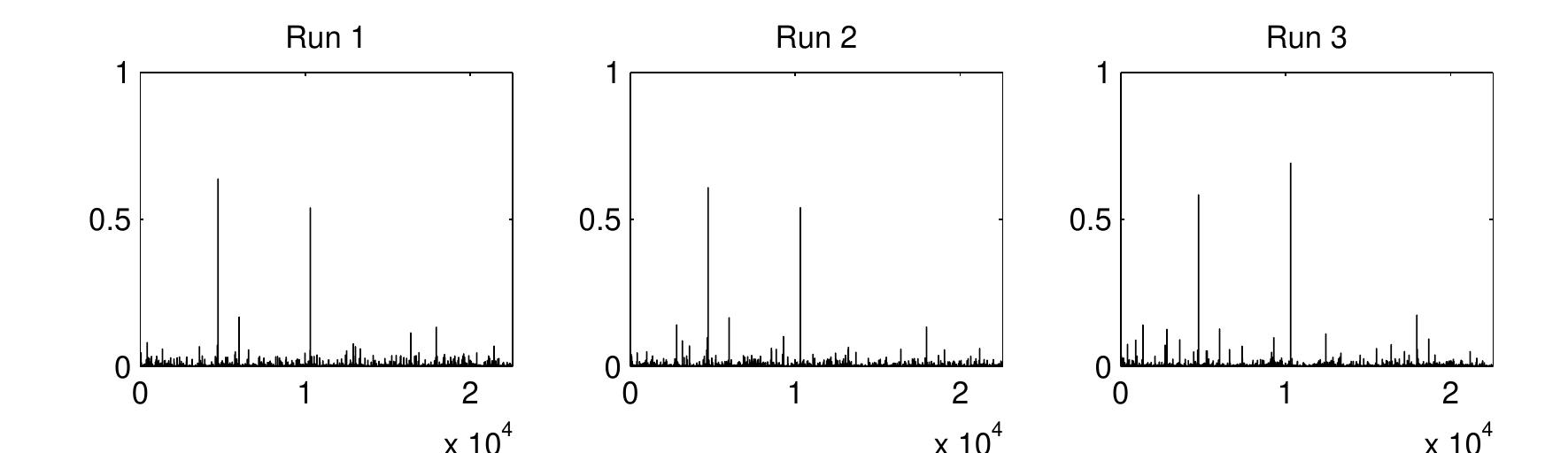}
\end{center}
\caption{\small PCR Data Example: PIP's for three runs of the MCA-IA-PT algorithm with
 6 temperatures}\label{gene1}
\end{figure}
 We consider one of these phenotypes, phosphoenopruvate carboxykinase (PEPCK) as the response variable. \cite{BoRe12}  apply their method to both a subset of 2\ 000 variables (selected on the basis of marginal correlations with the response) and the full data set. We use our adaptive algorithm on the full data set of 22\ 576 variables. In prior (\ref{prior}) we adopt $V_{\gamma}=100 I$ and a hierarchical  prior was used for $\gamma$ by assuming that $h\sim\Be(1, (p-5)/5)$ which implies that the prior mean number of included variables is 5. An MCA-PT-IA algorithm was run with
 $\tau=0.35$, $m=6$ temperatures, $r=5$ or $25$ multiple chains, and
24\ 000\ 000 iterations (the number of iterations for each chain was divided by the number of multiple chains leading to comparable computational times). 
Three independent runs of the algorithms were done for each combination of tuning parameters.

 Figure~\ref{gene1} shows the PIP's with 5 and 25 multiple chains. The results indicate that two genes are particularly predictive of the response with PIP's over 0.5. There are also many other variables with smaller but non-negligible PIP's.
\begin{figure}[h!]
\begin{center}
\begin{tabular}{c}
5 chains\\
\includegraphics[scale=0.8, clip, trim=10mm 0mm 20mm 245mm ]{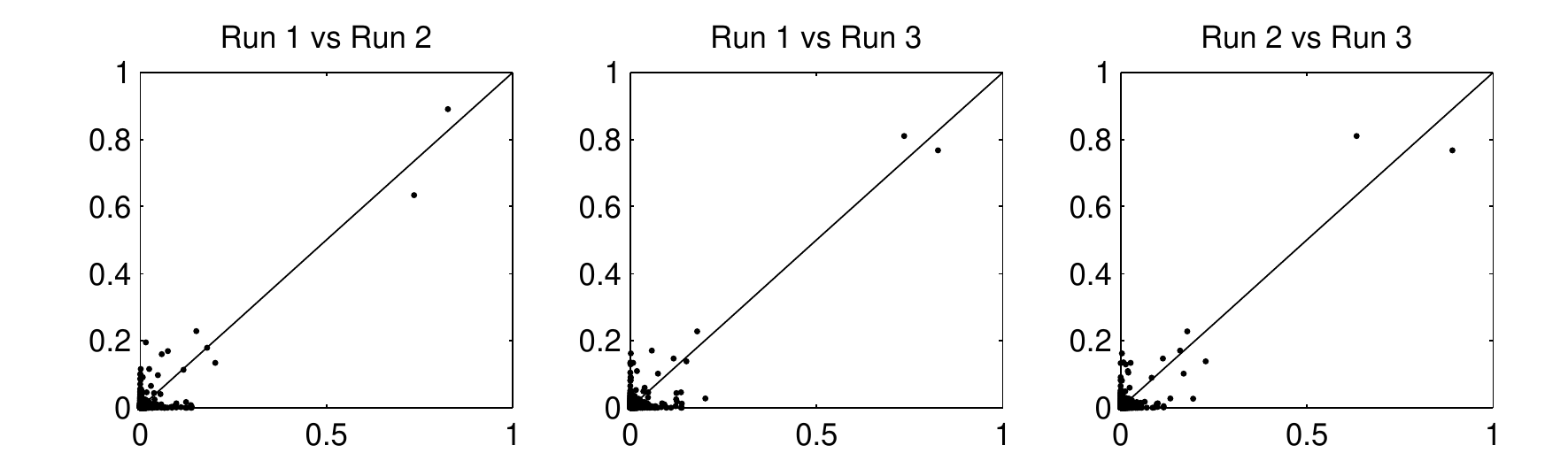}\\
25 chains\\
\includegraphics[scale=0.8, clip, trim=10mm 0mm 20mm 245mm ]{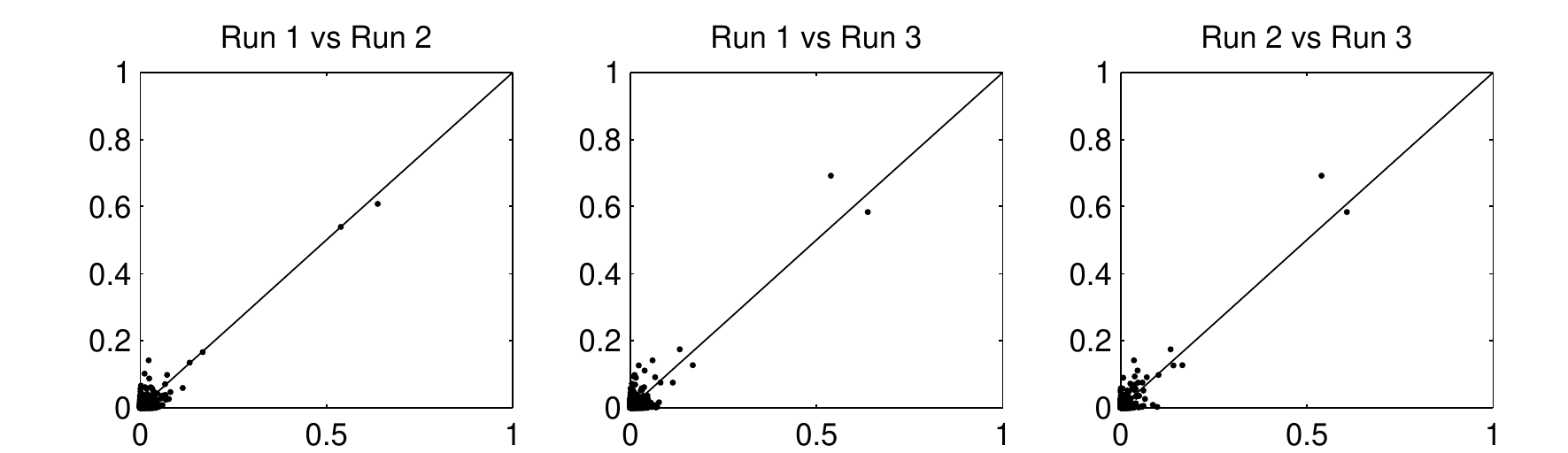}
\end{tabular}
\end{center}
\caption{\small PCR Data: scatter plots of the PIP's for three runs of the MCA-IA-PT algorithm with
 6 temperatures}\label{gene2}
\end{figure}
Results from the different runs are in good agreement. Figure~\ref{gene2} shows pairwise comparisons of the PIP's for each algorithmic parameter setting.
Estimated PIP's are close, particularly for the variables with high PIP's.
\begin{figure}[h!]
\begin{center}
\begin{tabular}{c}
5 chains\\
\includegraphics[scale=0.8, clip, trim=10mm 0mm 20mm 245mm ]{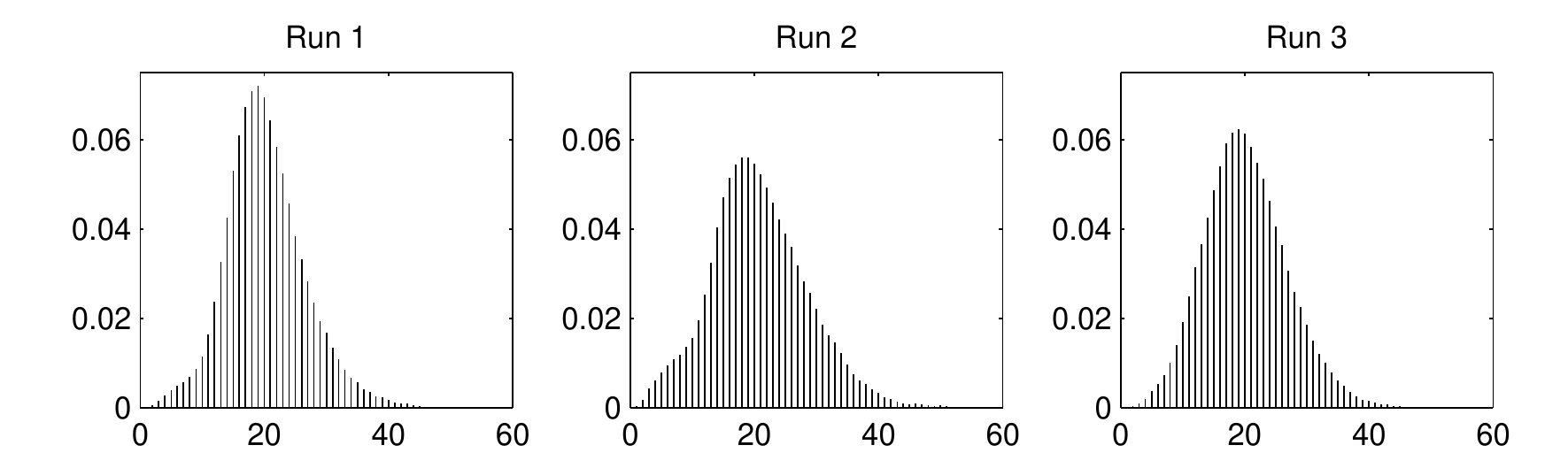}\\
25 chains\\
\includegraphics[scale=0.8, clip, trim=10mm 0mm 20mm 245mm ]{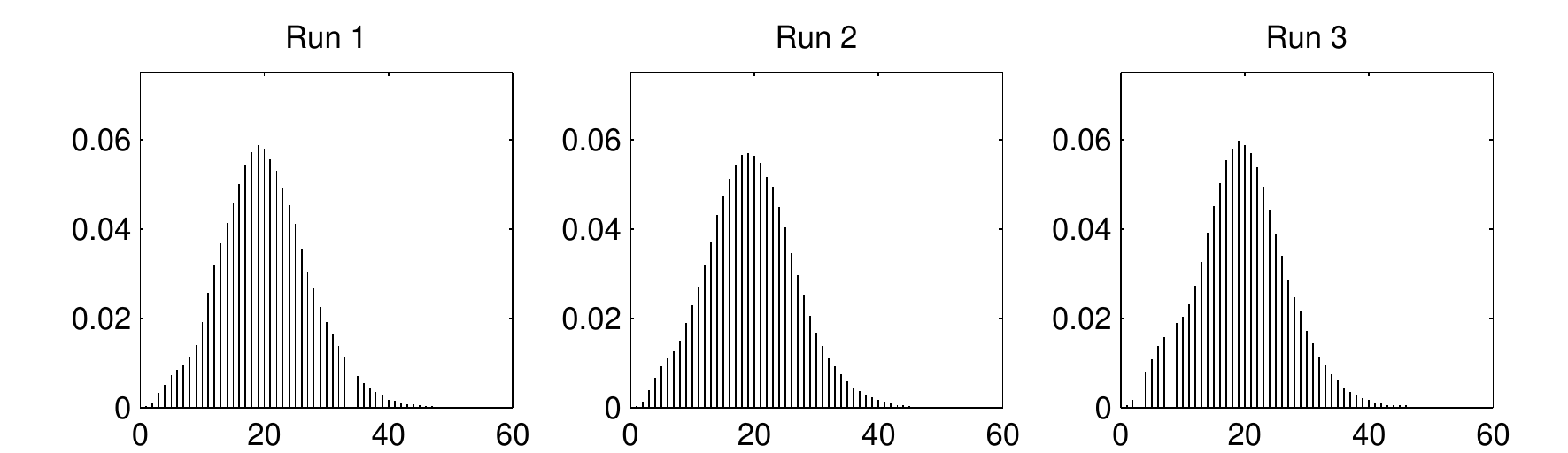}
\end{tabular}
\end{center}
\caption{\small PCR Data: posterior distribution of model size for three runs of the MCA-IA-PT algorithm with
 6 temperatures}\label{gene4}
\end{figure}
Figure~\ref{gene4} shows the posterior distribution of model size from the three runs.  The posterior mean model sizes calculated using output from the three runs were  20.2,  20.7 and 20.2 with 5 chains and
  20.2,  19.6 and   19.5 with 25 chains. This results is quite sensitive to the choice of the prior on model space. For example, setting $h=5/22\ 576$ (rather than using the hierarchical prior, while keeping the same prior mean model size) leads to much smaller model sizes.
 The posterior mean model sizes in the three runs were  8.8,  8.9 and 9.0 with 5 chains and
  8.4,  8.0 and  8.7 with 25 chains. This is in line with the fact that the prior with a fixed $h$ is much more informative than the hierarchical prior (see Ley and Steel, 2009\nocite{LeySteel09}). However, the ranking of the variables in terms of PIP is largely unchanged.
The posterior mean model size with the hierarchical prior is much larger than the ones reported by \cite{BoRe12} using their marginal sets method.

\begin{figure}[h!]
\begin{center}
\begin{tabular}{c}
5 chains\\
\includegraphics[scale=0.8, clip, trim=10mm 0mm 20mm 245mm ]{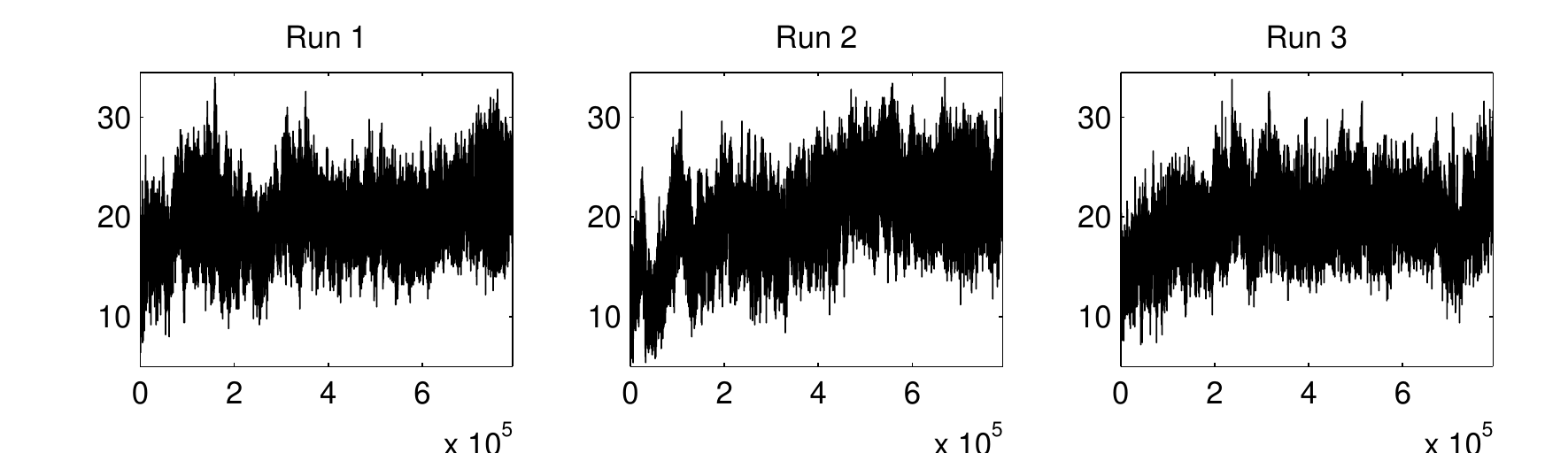}\\
25 chains\\
\includegraphics[scale=0.8, clip, trim=10mm 0mm 20mm 245mm ]{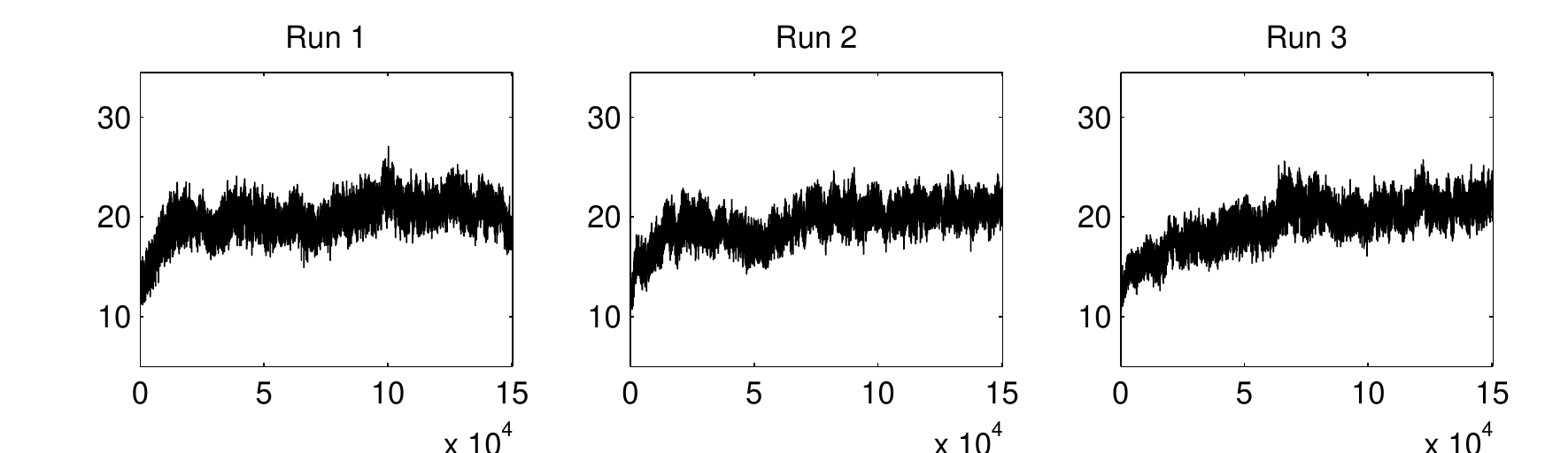}\\
\end{tabular}
\end{center}
\caption{\small PCR Data: trace plots of the model size averaged across the multiple chains using the MCA-IA-PT algorithm}\label{gene5}
\end{figure}
Figure~\ref{gene5} shows a trace plot of the model size averaged over the multiple chains. This indicate that the average model size for all runs stabilizes around 20. The results with 25 chains have a smaller variability since the average at every iteration involves a larger number of draws.

\cite{BoRe12} applied their method to a subset of 2\ 000 variables chosen to have the largest correlation with the response.
Figure~\ref{gene6} shows a scatter plot of the PIP's for these 2\ 000 genes with both the full data set and the subset.
Six of the eight genes with PIP's using the full data over 0.1 are included in the reduced data set (with the third and fourth most important genes being excluded). In addition, 10 of the 17 genes with PIP's over 0.05 are included and 43 of the 164 genes with PIP's over 0.01 are included. The diminishing proportion of genes included in the reduced data set as we lower the PIP threshold is not surprising since the reduced set is chosen using the marginal relationship between the response and the genes.
\begin{figure}[h!]
\begin{center}
5 chains\\
\includegraphics[scale=0.8, clip, trim=10mm 0mm 20mm 245mm ]{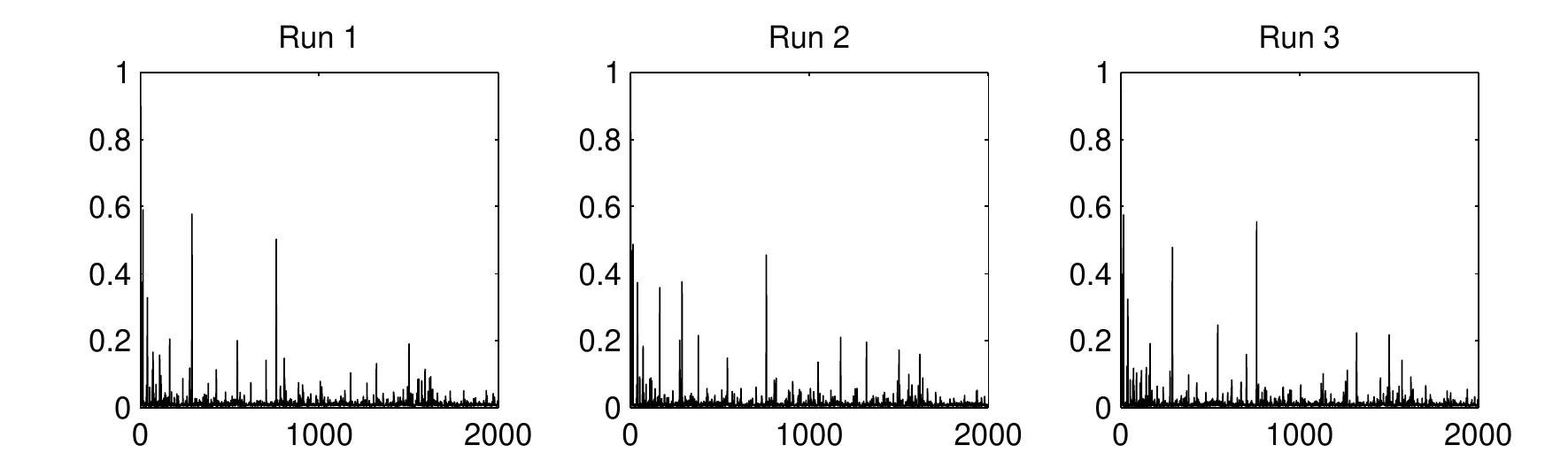}\\
25 chains\\
\includegraphics[scale=0.8, clip, trim=10mm 0mm 20mm 245mm ]{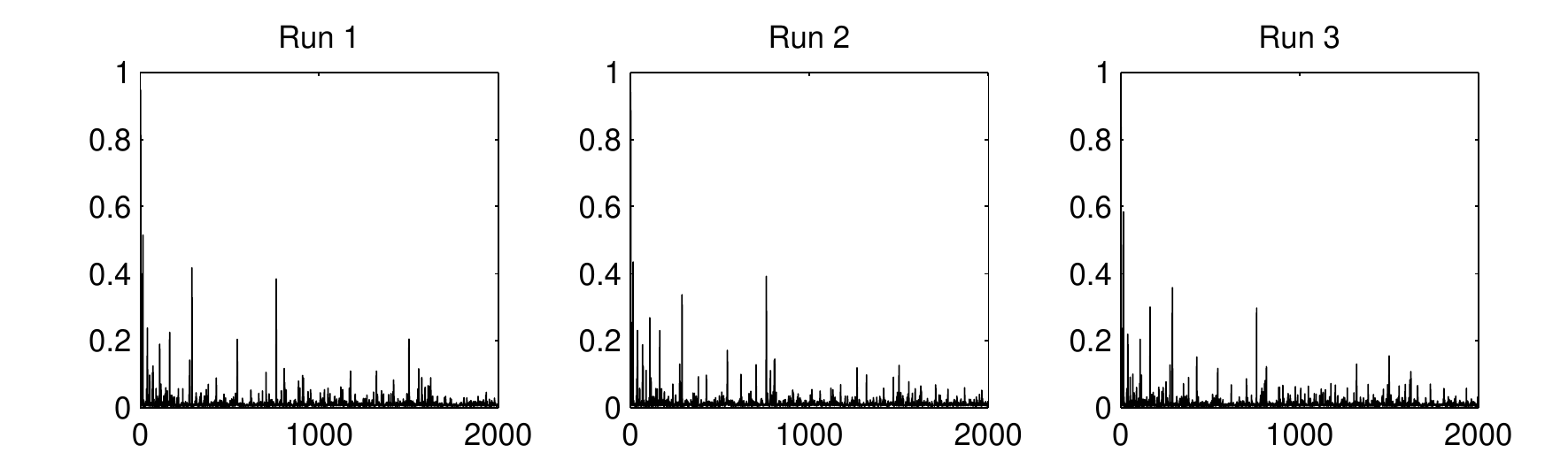}
\end{center}
\caption{\small PCR Data Example: PIP's for the reduced data using three runs of the MCA-PT-IA algorithm with
 6 temperatures and 5 chains}\label{gene3}
\end{figure}
Figure~\ref{gene3} shows the PIP's using only the reduced data set from three runs of the MCA-PT-IA algorithm with
 6 temperatures. The top two genes from the full data set are the most important but the discrimination between important and unimportant genes is less clear with several variables whose PIP's are around 0.5 using the subset but are much smaller using all the data (see Figure~\ref{gene4}).
The posterior mean model size with the reduced data set was 27.6 (averaged across the three runs) compared with 20.4 for the full data set. This suggests that the reduction method removes some simpler models which are well-supported by the data from the set of possible models. These results illustrate the potential problems that can arise by screening variables based on the marginal relationship with the response, such as the popular SIS (sure independence screening) and iterative SIS procedures of \cite{FanLv} and the Bayesian subset regression method of \cite{Liang_etal}. \cite{BoRe12} also use SIS on the full data set in combination with SCAD (smoothly clipped absolute deviation; Fan and Li, 2001\nocite{FanLi}) which results in very small models (mean model size is 2.3) and relatively poor prediction.

\begin{figure}[h!]
\begin{center}
\begin{tabular}{cc}
\includegraphics[scale=0.8, clip, trim=0mm 0mm 140mm 245mm ]{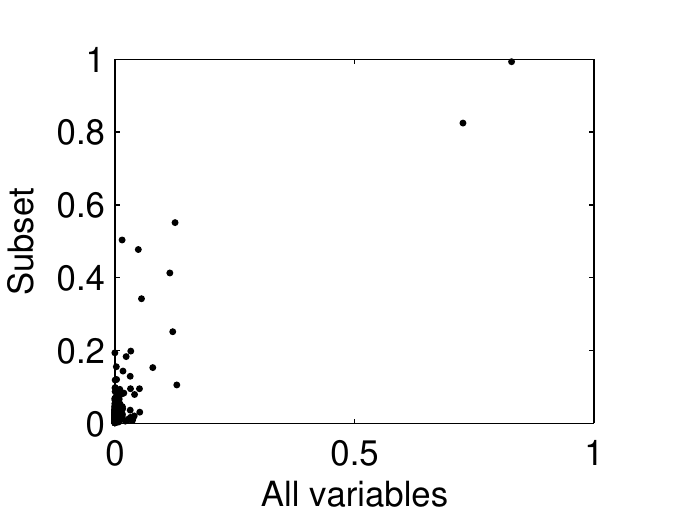}
\end{tabular}
\end{center}
\caption{\small PCR Data:
 A scatter plot of PIP's calculated using the full data set and the subset}\label{gene6}
\end{figure}

\section{Discussion}  \label{sec:concl}

Markov chain Monte Carlo methods for Bayesian variable selection has traditionally been considered a difficult problem associated with slow mixing. The individual adaptation algorithm is a method which can substantially improve mixing and lead to much more accurate estimates of posterior quantities, such as posterior inclusion probabilities. It leads to six- and seven-fold improvements in effective sample size in our examples and effectively opens the door for formal Bayesian model selection and model averaging analyses involving very large numbers of covariates, such as over 22 thousand in one of our examples.


These results illustrate the potential of carefully constructed adaptive Monte Carlo schemes in difficult problems for Bayesian inference. Much work on adaptive Monte Carlo has concentrated on problems where hand-tuning of algorithms is feasible but tiresome. The proposal in this paper has $2p$ parameters and tuning is only possible using adaptive Monte Carlo ideas. The development of similar algorithms where hand-tuning would be infeasible represents an interesting, and as yet virtually unexplored, area for future research.

\appendix
\section{Supplementary material for ``Individual adaptation: an adaptive MCMC scheme for variable selection problems''}

\subsection{Proofs of Ergodicity Results}

\begin{proof}[Proof of Lemma 1]

To verify the result  it is enough to check that the whole state space $M^{ r}$ is $1-$small with the same constant $\beta>0,$ (c.f. \cite{roberts2004general}), that is check for example that there exists $\beta > 0$ s.t. for every $\eta \in \Delta_{\varepsilon}$ and every $\gamma^{\otimes r}, \gamma'^{\otimes r} \in M^{ r}$ we have
\begin{equation}\label{eqn:Doeblin}
P_{\eta}(\gamma^{\otimes r}, \gamma'^{\otimes r}) \geq \beta.
\end{equation}
First decompose the move into proposal and acceptance
\[
P_{\eta}(\gamma^{\otimes r}, \gamma'^{\otimes r}) \; = \; q_{\eta}(\gamma^{\otimes r}, \gamma'^{\otimes r}) \times a_{\eta}(\gamma^{\otimes r}, \gamma'^{\otimes r}),
\]
and notice that by the proposal construction $q_{\eta}(\gamma^{\otimes r}, \gamma'^{\otimes r}) \geq \varepsilon^{rp}$ since $|M^r| = rp.$ Similarly
\begin{eqnarray*}
 a_{\eta}(\gamma^{\otimes r}, \gamma'^{\otimes r}) & = & \min\left\{1, {\pi^{\otimes r}(\gamma'^{\otimes r}) q_{\eta}(\gamma'^{\otimes r}, \gamma^{\otimes r}) \over \pi^{\otimes r}(\gamma^{\otimes r}) q_{\eta}(\gamma^{\otimes r}, \gamma'^{\otimes r}) }  \right\} \\ & \geq & \pi^{\otimes r}(\gamma'^{\otimes r}) q_{\eta}(\gamma'^{\otimes r}, \gamma^{\otimes r}) \; \geq \; \pi_{m}^r\times \varepsilon^{rp},
\end{eqnarray*}
where $\pi_{m} := \min_{\gamma \in M} \pi(\gamma).$ Consequently in \eqref{eqn:Doeblin} we can take \[\beta = \pi_{m}^r\times \varepsilon^{2rp},\] and we have established simultaneous uniform ergodicity.
\end{proof}

\begin{proof}[Proof of Theorem 1]
Theorem 1 follows from Theorem 1 (ergodicity) and Theorem 5 (WLLN) of
\cite{MR2340211}. Precisely, simultaneous uniform ergodicity for
nonadaptive kernels holds via Lemma 1. Moreover, it is routine to
check that the proposal satisfies diminishing adaptation, and
consequently by Lemma 4.21 (ii) of \cite{latuszynski2013adaptive}
applied with discrete topology of the variable selection context, also the
transition kernels satisfy diminishing adaptation {\it i.e.}~the random variable
\[
\mathcal{D}_i := \sup_{\gamma^{\otimes r} \in M^{ r}}\|P_{\eta^{(i+1)}}( \gamma^{\otimes r}, \cdot) - P_{\eta^{(i)}}( \gamma^{\otimes r}, \cdot)\|
\]
converges to $0$ in probability as $i \to \infty.$
\end{proof}

\begin{proof}[Proof of Theorem 2]

We conclude Theorem 2 from Theorem 1 (ergodicity) and Theorem 3 (WLLN)
of   \cite{ArIk13}. To this end we need an analogue of Lemma 1 for the
parallel tempering version of the kernel to verify
simultaneous uniform ergodicity. This can be established along the
same lines as Lemma 1, necessarily with additional notational
complication that we omit
here for brevity. Similarly, it is routine to verify that the parallel tempering adaptive kernel
proposals satisfy
diminishing adaptation and again by Lemma 4.21 (ii) of \cite{latuszynski2013adaptive}
applied with discrete topology of the variable selection context, also the
transition kernels satisfy diminishing adaptation.
\end{proof}

\subsection{Example: Boston Housing data}

We considered the Boston housing data previously analyzed by \cite{ScCh11} in the context of mixing of MCMC algorithms for variable selection. Here we have $n=506$ observations on the log of the median values of owner-occupied housing which are modelled through a linear regression model using $p=104$ potential covariates. We use the prior in equation (1) of the paper with $V_\gamma=100I$ and $h=5/104$. The problem differs from the previous one in that $n>p$, but there is reported evidence of multimodality in the posterior on model space. Thus, we use the methods described in Section 4
and consider the ability of our adaptive algorithm in combination with both the sequential Monte Carlo (SMC) and parallel tempering (PT) algorithms. The complicated nature of the posterior distribution is illustrated by the results in Table~\ref{BH_prob}. The two models with the largest posterior probability differ by only one variable. However, the difference between those models and the model with the third largest posterior probability is much greater. Therefore, it will be difficult for many MCMC algorithms to traverse this posterior distribution. The IA-SMC algorithm was run with 92\ 500 particles and $K=1$, 18\ 500 particles and $K=5$, 9\ 250 particles and $K=10$,
3\ 700 particles and $K=25$ and finally 1\ 850 particles and $K=50$ and the IA-PT algorithm ($m=8$) was run with a burn-in period of 12\ 500 and, subsequently, for  525\ 000 iterations
with no  thinning. This was found to lead to similar run-times for the different algorithms.

\begin{sidewaystable}
\centering
\begin{tabular}{ccccccccccccccccccccccccccc}
& & &  & & & & & & & & & & & & & & & & & & & & & & & Post. Prob.\\
5  &   6  &   8 &       &        &       &       & 13   &       & 24   & 29  &  49  &       &       & 55 &   58 &         &       & 67 &  78 &   86 &   91 &   97  & 101 &        &        & 0.243\\
 5  &   6  &   8 &       &        &       &       & 13   &       & 24   & 29  &        &       &       & 55 &   58  &        &      & 67  &  78 &   86  &  91 &   97  & 101 &        &        & 0.140\\
 5  &   6  &   8 &       &        &  11 &  12 & 13   &       &        & 29  &  49  &  50 &       & 55 &         & 59  &  61 &       &       &   86  &       &  97   & 101 &        &        & 0.031\\
 5  &   6  &   8 &       &        &       &       & 13   &       & 24  &  29  &  49  &       &       & 55 &   58   &      &       & 67   & 78 &   86  &  91 &   97  & 101 & 103 &        & 0.022\\
 5  &   6  &   8 &   9  &  10  &       &       & 13   &       & 24  &  29  &  49  &       &       & 55 &         &       &       &        & 78 &   86  &  91 &   97  & 101 &        &        & 0.021\\
 5  &   6  &   8 &       &        &       &       & 13   &       &       & 29   &  49  &  50 &       & 55 &         & 59  &        &        &78 &   86  &  91  &  97  & 101 &        &        & 0.018\\
 5  &   6  &   8 &       &        &       &       & 13   &       &       & 29   &  49  &  50 &       & 55 &   58  & 59 &       &         &78 &   86  &  91 &   97  & 101 &        &        & 0.016\\
 5  &   6  &   8 &       &        &       &       & 13   &  14 &       &        &  49  &  50  &      & 55 &         & 59 &        &         & 78 &  86  &  91 &   97  & 101 &        &        & 0.015\\
 5  &   6  &   8 &       &        &       &       & 13   &       &       & 29   &  49  & 50   & 54 & 55  &        & 59 &        &         &78   & 86  &  91 &   97  & 101 &        &        & 0.013\\
 5  &   6  &   8 &       &        &       &       & 13   &       & 24  &  29  &        &        &      & 55 &   58  &      &        & 67   & 78  &  86  &  91 &   97  & 101 &        &104 &  0.013\\
 5  &   6  &   8 &       &        &       &       & 13   &  14 & 24  &        & 49   & 50   &      & 55  &        & 59 &        &        &78  &   86  &  91  &  97  & 101 &        &       & 0.012\\
 5  &   6  &   8 &   9  &  10  &       &       & 13   &       & 24  & 29   &        &        &      & 55  &        &      &        &         &78 &   86  &  91   & 97  & 101 &        &       &  0.011\\
 5  &   6  &   8 &       &        &       &       & 13   &       &       & 29   &  49  &  50  &      & 55  &        & 59 &        &67    & 78  &  86  &  91 &   97  & 101 &        &       & 0.011\\
\end{tabular}
\caption{The 10 models with the highest posterior probability for the Boston housing data (the variable names are given in the Appendix).}\label{BH_prob}
\end{sidewaystable}

\begin{figure}[h!]
\begin{center}
\includegraphics[trim=0mm 60mm 10mm 90mm, scale=0.5, clip]{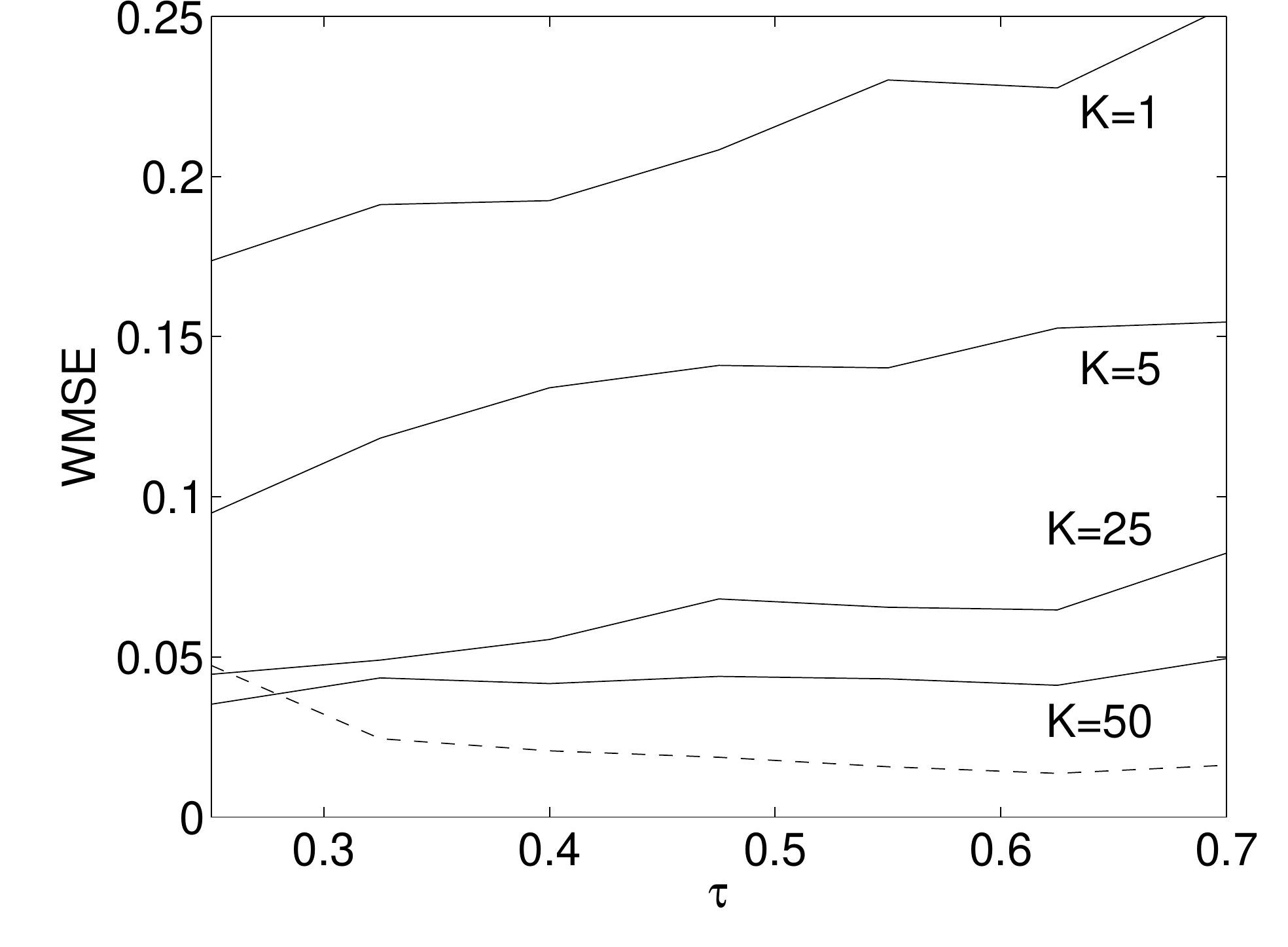}
\end{center}
\caption{\small Boston Housing data: The Weighted Mean Squared Error using IA-SMC (solid line) with $K=1$, $K=5$, $K=25$  and $K=50$ and the IA-PT algorithm (dashed line).}\label{BH_ESS}
\end{figure}
The ESS may not be well-estimated from a single run if the run leans to biased estimates. An alternative is the mean squared error of the estimate across multiple runs. This is an estimate of the variance of the Monte Carlo estimate when the Monte Carlo estimates are unbiased. However, it naturally includes a penalty for the sampler producing biased estimates.
Rather than use Mean Squared Error, the accuracy of the algorithms was evaluated using a Weighted Mean Squared Error (WMSE) 
\[
\mbox{WMSE}=\sum_{i=1}^M \sum_{j=1}^p w_j(\hat\theta_{ij}-\theta^{\star}_j)^2
\]
where $M$ is the number of replicate MCMC or SMC runs,  $\hat\theta_{ij}$ is the estimated posterior inclusion probability for the $j$-th variables in the $i$-th run and $\theta^{\star}$ is a ``gold-standard'' estimate of the posterior inclusion probability for the $j$-th variable. The weights $w_j$ are assumed to be such that $\sum_{j=1}^p w_j=1$ and $w_j$ represents the importance of the $j$-th variables. We chose $M=60$ and
$w_j\propto\theta^{\star}_j$ in our comparisons. The gold standard value of $\theta^{\star}_j$ was calculated using output from the PT chains and SMC with $K=25$ and $K=50$ which had the highest levels of accuracy.

The WMSE  is shown in Figure~\ref{BH_ESS}. The WMSE for the PT algorithm with $m=8$ is shown as a dashed line and decreases with $\tau$. The graph also shows the WMSE's for the SMC algorithm with $K$ MCMC steps in the re-weighting step.
These range from $K=1$ to $K=50$. The WMSE decreases with the number of steps for each value of $\tau$ with the WMSE for $K=50$ having a similar WMSE to the PT algorithm for small $\tau$ but for $\tau\ge 0.3$  the PT algorithm does a lot better.
 The effect of $\tau$ on the WMSE differs according to the number of Metropolis-Hastings steps. The WMSE tends to increase with $\tau$ for $K=1$ and $K=5$ whereas WMSE is not that much affected by $\tau$ for $K=10$, $K=25$ and $K=50$.
The simple Metropolis-Hastings algorithm was run with a burn-in period of $12 500$ with
9\ 750\ 000 subsequent iterations with no thinning. This took the same computational times as the other algorithms and so represents a comparison to the more complicated algorithms for multi-modal distributions. The WMSE for the simple MH algorithm was 0.0071 which is smaller than all algorithm apart from the IA-SMC algorithm with $K=50$ and $K=25$ with smaller values of $\tau$ and the IA-PT algorithm. The improvement of the IA-PT over the simple MH algorithm is still substantial. The acceptance rate is roughly 2\% for the simple MH algorithm and so the adaptive algorithm of \cite{lamnisos12} would reduce to the simple MH algorithm for this data set.



\bibliographystyle{Chicago}
\bibliography{References2}

\appendix
\subsection*{Appendix: Variables for the Boston housing data}

This is a list of the variables that appear in Table~\ref{BH_prob} using the names given in the R package \verb+spdep+.\\
\begin{center}
\begin{tabular}{cccccc}
5 & NOX & 24 & NOX $\times$ CRIM &  67 & TAX $\times$ RAD\\
6 & RM & 29 & RM $\times$ CRIM & 78 & PTRATIO $\times$ TAX\\
8 & DIS & 49 & $\mbox{DIS}^2$ & 86 & B $\times$ DIS\\
9 & RAD & 50 & RAD $\times$ CRIM & 91 & $\mbox{B}^2$\\
10 & TAX & 54 & RAD $\times$ NOX & 97 & LSTAT $\times$ RM\\
11 & PTRATIO & 55 & RAD $\times$ RM & 101 & LSTAT $\times$ TAX\\
12 & B & 58 & $\mbox{RAD}^2$ & 103 & LSTAT $\times$ B\\
13 & LSTAT & 59 & TAX $\times$ CRIM & 104 & $\mbox{LSTAT}^2$\\
14 & $\mbox{CRIM}^2$ & 61 & TAX $\times$ CHAS &
\end{tabular}
\end{center}

\bibliographystyle{Chicago}
\bibliography{References2}

\end{document}